\documentclass[a4paper,fleqn,usenatbib,useAMS]{mnras}

\usepackage{graphicx}   
\usepackage{amsmath}    
\usepackage{amssymb}    
\usepackage{multicol}        
\usepackage{bm}         
\usepackage{pdflscape}  

\usepackage[T1]{fontenc}
\usepackage{ae,aecompl}

\title[Investigating the counterparts to unidentified sources in the 1000-orbit \emph{INTEGRAL}/IBIS catalogue]
{Investigating the X-ray counterparts to unidentified sources in the 1000-orbit \emph{INTEGRAL}/IBIS catalogue}

\author[R. Landi]{R. Landi$^{1}$\thanks{Contact e-mail:
landi@iasfbo.inaf.it}, L.~Bassani$^{1}$, A.~Bazzano$^{2}$, A. J. Bird$^{3}$, M.~Fiocchi$^{2}$,
A.~Malizia$^{1}$, \newauthor F.~Panessa$^{2}$, V.~Sguera$^{1}$, P.~Ubertini$^{2}$\\   
$^{1}$ INAF -- IASF Bologna, Via P. Gobetti 101, I--40129 Bologna, Italy\\
$^{2}$ INAF -- IAPS Rome, Via Fosso del Cavaliere 100, I--00133 Roma, Italy\\
$^{3}$ School of Physics and Astronomy, University of Southampton, SO17 1BJ, Southampton, UK}
\date{Last updated 2015 May 22; in original form 2013 September 5}

\pubyear{2015}

\date{Last updated 2015 May 22; in original form 2013 September 5}

\pubyear{2015}

\begin{document}
\label{firstpage}
\pagerange{\pageref{firstpage}--\pageref{lastpage}}
\maketitle

\begin{abstract}

The latest \emph{INTEGRAL}/IBIS all-sky survey lists 219 hard X-ray sources whose nature is 
still unknown. We report on our ongoing campaign aimed at identifying these high-energy 
emitters by exploiting the focusing capabilities of the X-ray Telescope (XRT, 0.2--10 keV) on 
board \emph{Swift}, which allow an enhancement of the source localisation to 
arcsec level, thus facilitating the identification of the likely counterpart. By 
cross-correlating the list of the unidentified IBIS sources included in the latest IBIS 
catalogue with \emph{Swift}/XRT archival data, we found a set of 14 objects, not yet reported 
in the literature, for which XRT data were available. We found no detection in only one case, a 
single X-ray association in 9 sources, and 2/3 associations in the remaining objects. We then 
made use of multi-waveband archives to search for counterparts at other wavelengths of these 
XRT detections and exploited X-ray spectral information in an attempt to determine their 
nature and association with the IBIS object. As a result of our analysis, we identified a 
single counterpart for 13 sources, although in some cases its nature/class could not be 
assessed on the basis of the information collected. More specifically, we found that 
SWIFT J0924.2$-$3141 and SWIFT J1839.1$-$5717 are absorbed AGN, while SWIFT J0800.7$-$4309 and 
1SWXRT J230642.8$+$550817 are Cataclysmic Variable binary systems. Finally, we found that IGR 
J14059$-$6116 is likely associated with the \emph{Fermi} source 3FGL J1405.4$-$6119. In the 
case of XMMSL1 J030715.5$-$545536 no XRT counterpart was detected. 
In all the other cases, 
optical/infrared spectroscopy is necessary to classify properly each X-ray counterpart and 
confirm their association with the \emph{INTEGRAL}/IBIS detection.

\end{abstract}





\section{Introduction}
Over the last decade, our knowledge of the soft gamma-ray sky ($>$10 keV) has been 
significantly revolutionised thanks to the results obtained by IBIS (Ubertini et al. 2003) on 
board \emph{INTEGRAL} (Winkler et al. 2003) and the Burst Alert Telescope (BAT, Barthelmy et 
al. 2005) on board \emph{Swift} (Gehrels et al. 2004). Both telescopes operate in similar 
wavebands (around 20--200 keV) with a limiting sensitivity of about a mCrab and a point source 
location accuracy of the order of a few arcmin, depending on the source strength. Both 
instruments continue to survey the sky at high energies, thus providing an unprecedented 
sample of objects selected in the soft gamma-ray band. A significant fraction of these sources 
are still unidentified/unclassified, often because they lack coverage in the 2--10 keV energy 
range. The recent all sky \emph{INTEGRAL}/IBIS survey (Bird et al. 2016) lists 939 soft 
gamma-ray selected sources of which 219 are still unassociated/unidentified. The 
identification process is crucial if one wants to gain an insight into the nature of the 
sources that populate our Universe at soft gamma-ray energies. To this aim, a refined 
localisation, atteinable by exploiting the capability of current focusing X-ray telescopes, is 
necessary to pinpoint and classify their optical counterparts. Furthermore, information in the 
X-ray band can help to characterise these sources in terms of spectral shape, flux, absorption 
properties and variability.

In this paper, we present the results of our ongoing campaign focused on identifying the still 
unknown \emph{INTEGRAL}/IBIS sources. To this aim, we searched for X-ray archival data 
acquired with XRT on board the \emph{Swift} satellite available before the end of June 2016, 
finding a set of 14 objects for which low energy data can provide X-ray information. Most of 
these sources are on the Galactic plane except for five objects (XMMSL1 J030715.5$-$545536, 
IGR J0924.2$-$3141, 1RXS J145959.4$+$120124, IGR J18074$+$3827, and SWIFT J1839.1$-$5717), 
which are instead located at high (above 20 degrees) Galactic latitudes.

The paper is structured as follows: in Sect. 2 we present the XRT data reduction and analysis 
and the criteria adopted to search for the likely counterparts to the IBIS sources. Sect. 3 is 
devoted to the discussion of the results for each individual source. Conclusions are drawn in 
Sect. 4.

\section{\emph{Swift}/XRT data reduction and analysis}

The log of all X-ray observations taken into account in this work is shown in Table~\ref{tab1} 
where we report for each individual IBIS source, the observation ID, as well as the date and 
the on-source exposure time of each XRT pointing. The XRT data of the 14 selected sources were 
reduced by means of the XRTDAS standard data pipeline package ({\sc xrtpipeline} v. 0.13.2) to 
produce screened event files. All data were extracted only in the Photon Counting (PC) mode 
that ensures a source fine positioning.

We then, for each IBIS source, summed together all the available XRT pointings using {\sc 
XSELECT} v. 2.4c to enhance the signal-to-noise ratio and thus facilitate the detection of 
candidate counterparts. As a following step, we analysed the XRT images in the 0.3--10 keV 
energy band by means of {\sc XIMAGE} v. 4.5.1 in search of X-ray detections above 3$\sigma$ 
confidence level (c.l.) within both the 90\% and 99\% IBIS error circles. Furthermore, we 
checked the XRT images in the 3--10 keV band to select those sources with the hardest spectra 
(i.e. those with detection above 3 keV), since these are most likely to be the counterparts to 
the IBIS objects. For this reason, throughout the paper we will restrict our discussion to 
these sources, providing details on weaker X-ray detections only when possibly relevant. We 
estimated the X-ray positions using the task {\sc xrtcentroid v.0.2.9}. In the XRT images we 
plot the 90\% and 99\% IBIS positional uncertainties (black and black-dotted circles, 
respectively) and the 90\% \emph{Swift}/BAT error circle (black-dashed-dotted circle) when 
available. To visualise better the X-ray counterparts, in most cases we smoothed the images. 
Therefore, the presence of grains and/or features inside the XRT field of view are undoubtedly 
spurious; in some cases the poor quality of the XRT images is due to the low exposure.

For spectral analysis, source events were extracted within a circular region with a radius of 
20 pixels (1 pixel $\sim$2.36 arcsec) centred on the source position, while background events 
were extracted from a source-free region close to the X-ray source of interest. The 
spectra were obtained from the corresponding event files using the {\sc XSELECT} v. 2.4c 
software and generally binned using {\sc grppha} to 20 counts per energy bin so that the 
$\chi^{2}$ statistic could be applied. For sources with fewer counts (typically around 
50--60), data were binned to 1 count per energy bin and the Cash statistic (Cash 1979) was 
adopted. We used version v.014 of the response matrices and created individual ancillary 
response files \emph{arf} using {\sc xrtmkarf} v.0.6.3.

For objects with more than one pointing, we first checked variability by analysing each single 
observation, then performed the spectral analysis of the average spectrum; this approach may 
provide further information that could be used as a possible filter on counterparts. For the 
spectral analysis, in the first instance, we adopted a basic model consisting of a simple 
power law passing through Galactic absorption in the source direction (Kalberla et al. 2005). 
If this baseline model was not adequate to fit the data, we then introduced extra spectral 
components as required.

\begin{table*}
\centering
\caption{Log of the \emph{Swift}/XRT observations used in this paper.}
\label{tab1}
\begin{tabular}{lccc}
\hline
\hline
IBIS source & ID & Obs. Date & Exposure$^{a}$ \\
       &          &      & (s) \\
\hline
\hline
XMMSL1 J030715.5$-$545536  & 00034389001  & Mar 03, 2016 &  3277 \\
                           & 00034389002  & Mar 08, 2016 &  2314 \\
\hline
SWIFT J0800.7$-$4309       & 00041761001  & Dec 08, 2010 &   692 \\
                           & 00041761002  & Dec 12, 2010 &  5509 \\
                           & 00041761003  & Dec 15, 2010 &  5717 \\
                           & 00041761004  & Dec 16, 2010 &  1063 \\
\hline
SWIFT J0924.2$-$3141       & 00091688001  & Apr 02, 2013 &  2600 \\
                           & 00091688002  & Apr 07  2013 &  2352 \\
                           & 00080674001  & Apr 19, 2014 &  6622 \\
\hline
IGR J14059$-$6116          & 00041805001  & Oct 04, 2011 &  3114  \\
                           & 00041805005  & Sep 21, 2012 &  4623  \\
\hline
1RXSJ 145959.4$+$120124    & 00034408001  & Mar 15, 2016 &   577  \\ 
                           & 00034408002  & Mar 16, 2016 &   303  \\
                           & 00034408003  & Mar 17, 2016 &   399  \\
                           & 00034408004  & Mar 18, 2016 &   652  \\ 
\hline
IGR J15038$-$6021          & 00046303001  & May 12, 2013 &   331 \\
                           & 00046303002  & Jul 04, 2013 &  1299 \\
\hline
IGR J16447$-$5138          & 00034390001  & Mar 04, 2016 &  2695 \\
                           & 00034390002  & Mar 06, 2016 &  1850  \\
\hline
IGR J17508$-$3219          & 00034409001  & Mar 18, 2016 &  3731 \\
\hline
IGR J18007$-$4146          & 00085658001  & Feb 06, 2016 &   233  \\
                           & 00085658002  & Mar 20, 2016 &  3016  \\
\hline
IGR J18074$+$3827          & 00034391001  & Mar 06, 2016 &  4207  \\
\hline
XMMSL1 J182831.8$-$022901  & 00049376001  & Nov 09, 2012 &   148   \\ 
                           & 00049376003  & Feb 09, 2013 &   421  \\
                           & 00049376004  & May 23, 2013 &   401  \\ 
                           & 00049376005  & May 30, 2013 &   181  \\
                           & 00049376006  & Jun 13, 2013 &   108  \\
                           & 00049376007  & Jun 19, 2013 &   266  \\
                           & 00049376008  & Jun 21, 2013 &  3741  \\
\hline
SWIFT J1839.1$-$5717       & 00038080001  & Oct 17, 2008 &  7675  \\
                           & 00038080002  & Nov 02, 2008 &  8364 \\                                            
\hline
IGR J20310$+$3835          & 00049381001  & Dec 04, 2012 &   108 \\
                           & 00049381002  & Dec 05, 2012 &  1028 \\
                           & 00049381003  & Dec 06, 2012 &  1550 \\
                           & 00049381004  & Dec 07, 2012 &   281 \\
                           & 00049381005  & Dec 10, 2012 &   672 \\
                           & 00049381006  & Dec 11, 2012 &   135 \\
                           & 00049381007  & Dec 12, 2012 &  1053 \\
\hline
1SWXRT J230642.8$+$550817  & 00039882001  & Sep 01, 2010 &   592  \\ 	
                           & 00039882002  & Oct 26, 2010 &   354  \\
\hline
\hline
\end{tabular}
\begin{list}{}{}
\item $^{a}$ On-source exposure time.  
\end{list}
\end{table*}

\begin{table*}
\footnotesize
\begin{center}
\footnotesize
\caption{\emph{INTEGRAL}/IBIS position of the 14 selected sources 
and locations of the objects detected by XRT, within the 90\% and 99\% IBIS positional uncertainties, 
with relative count rates in the 0.3--10 and 3--10 keV energy range, the number of X-ray observations
analysed and the total on-source exposure time.
The XRT error radii are given at 90\% confidence level.} 
\label{tab2}
\scriptsize
\begin{tabular}{lcccccc}
\hline
\hline
XRT source  &     R.A.     &     Dec.   &   error   &  \multicolumn{2}{c}{Count rate} & N. obs/Total expo    \\
            &              &            &          &   (0.3--10 keV)  &  (3--10 keV) &  \\   
  &   (J2000) &  (J2000) &   (arcsec)  &  (10$^{-3}$ counts s$^{-1}$) & (10$^{-3}$ counts s$^{-1}$) & (s)    \\
\hline
\hline 
\multicolumn{7}{c}{\textbf{XMMSL1 J030715.5$-$545536 (R.A.(J2000) = $03^{\rm h}06^{\rm m}55^{\rm s}.20$,
Dec.(J2000) = $-$$54^\circ54^{\prime}28^{\prime \prime}.80$, error radius (90\%)= 5$^{\prime}$.04;
(99\%) = 7$^{\prime}$.86)}}\\
   & &     &   &   &  &   \\
\multicolumn{6}{c}{no detection} & 2/5591 \\
\hline
\multicolumn{7}{c}{\textbf{SWIFT J0800.7$-$4309 (R.A.(J2000) = $08^{\rm h}00^{\rm m}22^{\rm s}.08$, Dec.(J2000) =
$-$$43^\circ09^{\prime}57^{\prime \prime}.60$, error radius (90\%)= 4$^{\prime}$.86;
(99\%) = 7$^{\prime}$.58)}}\\
   & &     &   &   &  &   \\
\#1 (in 90\%) & $08^{\rm h}00^{\rm m}40^{\rm s}.18$ & $-$$43^\circ11^{\prime}07^{\prime \prime}.27$ & 3.62 &
$63.80\pm2.90$ &  $30.30\pm2.00$ &  4/12981  \\
\#2 (in 90\%) & $08^{\rm h}00^{\rm m}45^{\rm s}.85$ & $-$$43^\circ09^{\prime}37^{\prime \prime}.53$ & 4.32 &
$3.03\pm0.74$  &  $1.29\pm0.38$   &   \\
\#3 (in 90\%) & $08^{\rm h}00^{\rm m}21^{\rm s}.75$ & $-$$43^\circ10^{\prime}41^{\prime \prime}.96$ & 4.60 &
$3.86\pm0.81$ &  --   &   \\
\hline
\multicolumn{7}{c}{\textbf{SWIFT J0924.2$-$3141 (R.A.(J2000) = $09^{\rm h}23^{\rm m}52^{\rm s}.56$,
Dec.(J2000) = $-$$31^\circ42^{\prime}21^{\prime \prime}.60$, error radius (90\%) = 4$^{\prime}$.32;
(99\%) = 6$^{\prime}$.74)}}\\
   & &     &   &   &  &   \\
\#1 (in 90\%) & $09^{\rm h}23^{\rm m}53^{\rm s}.61$ & $-$$31^\circ41^{\prime}31^{\prime \prime}.56$ & 4.11 &
$5.36\pm0.94$  &  $5.04\pm0.89$ &  3/11574  \\
\#2 (in 99\%) & $09^{\rm h}24^{\rm m}18^{\rm s}.18$ & $-$$31^\circ42^{\prime}17^{\prime \prime}.91$ & 3.51 &
$1885.0\pm15.0$  &  $599.5\pm8.3$   &   \\
\hline
\multicolumn{7}{c}{\textbf{IGR J14059$-$6116 (R.A.(J2000) = $14^{\rm h}05^{\rm m}56^{\rm s}.40$, Dec.(J2000) =
$-$$61^\circ16^{\prime}30^{\prime \prime}.00$, error radius (90\%) = 4$^{\prime}$.13; 
(99\%) = 6$^{\prime}$.44)}}\\
   & &     &   &   &  &   \\
\#1 (in 99\%) & $14^{\rm h}05^{\rm m}13^{\rm s}.93$ & $-$$61^\circ18^{\prime}29^{\prime \prime}.62$ & 5.24 &
$4.34\pm1.10$ &  $3.26\pm0.96$  & 2/7737  \\
\hline
\multicolumn{7}{c}{\textbf{1RXSJ 145959.4$+$120124 (R.A.(J2000) = $14^{\rm h}59^{\rm m}57^{\rm s}.84$,
Dec.(J2000) = $+$$12^\circ00^{\prime}10^{\prime \prime}.80$, error radius (90\%) = 4$^{\prime}$.39;
(99\%) = 6$^{\prime}$.85)}}\\
   & &     &   &   &  &   \\
\#1 (in 90\%)  & $14^{\rm h}59^{\rm m}59^{\rm s}.15$ & $+$$12^\circ01^{\prime}21^{\prime \prime}.94$ & 3.90 &
$120.4\pm10.0$ &  $21.55\pm4.40$  & 4/1931    \\
\hline
\multicolumn{7}{c}{\textbf{IGR J15038$-$6021 (R.A.(J2000) = $15^{\rm h}03^{\rm m}45^{\rm s}.84$, Dec.(J2000) =
$-$$60^\circ21^{\prime}25^{\prime \prime}.20$, error radius (90\%) = 3$^{\prime}$.95;
(99\%) = 6$^{\prime}$.16)}}\\
   & &     &   &   &  &   \\
\#1 (in90\%)  & $15^{\rm h}04^{\rm m}15^{\rm s}.99$ & $-$$60^\circ21^{\prime}21^{\prime \prime}.62$ & 5.11 &
$27.70\pm5.50$  & $11.77\pm3.50$  & 2/1630 \\
\hline
\multicolumn{7}{c}{\textbf{IGR J16447$-$5138 (R.A.(J2000) = $16^{\rm h}44^{\rm m}42^{\rm s}.72$,
Dec.(J2000) = $-$$51^\circ38^{\prime}56^{\prime \prime}.40$, error radius (90\%) = 4$^{\prime}$.95;
(99\%) = 7$^{\prime}$.72)}}\\
   & &     &   &   &  &   \\
\#1 (border of 90\%) & $16^{\rm h}44^{\rm m}32^{\rm s}.89$ & $-$$51^\circ34^{\prime}12^{\prime \prime}.63$ & 3.82 
& $71.03\pm5.30$ &  $19.92\pm2.80$ & 2/4545 \\
\hline
\multicolumn{7}{c}{\textbf{IGR J17508$-$3219 (R.A.(J2000) = $17^{\rm h}50^{\rm m}53^{\rm s}.04$,
Dec.(J2000) =  $-$$32^\circ19^{\prime}48^{\prime \prime}.00$, error radius (90\%) = 2$^{\prime}$.31;
(99\%) = 3$^{\prime}$.60)}}\\
   & &     &   &   &  &   \\
\#1 (in 90\%) & $17^{\rm h}50^{\rm m}55^{\rm s}.18$ & $-$$32^\circ18^{\prime}56^{\prime \prime}.07$ & 5.24 &
$8.25\pm2.20$ &  $4.64\pm1.50$  & 1/3731  \\
\#2 (in 99\%) & $17^{\rm h}51^{\rm m}06^{\rm s}.47$ & $-$$32^\circ18^{\prime}24^{\prime \prime}.13$ & 3.80 &
$86.04\pm6.40$ &  --  &   \\
\hline
\multicolumn{7}{c}{\textbf{IGR J18007$-$4146 (R.A.(J2000) = $18^{\rm h}00^{\rm m}48^{\rm s}.48$,
Dec.(J2000) = $-$$41^\circ48^{\prime}07^{\prime \prime}.20$, error radius (90\%) = 3$^{\prime}$.19;
(99\%) = 4$^{\prime}$.98)}}\\
   &    &    &    &    &    &     \\
\#1 (in 90\%) &  $18^{\rm h}00^{\rm m}42^{\rm s}.50$ & $-$$41^\circ46^{\prime}48^{\prime \prime}.75$ & 4.04 &
$49.42\pm5.20$ &  $16.51\pm3.00$  & 2/3249  \\
\hline
\multicolumn{7}{c}{\textbf{IGR J18074$+$3827 (R.A.(J2000) = $18^{\rm h}07^{\rm m}40^{\rm s}.80$,
Dec(J2000) = $+$$38^\circ26^{\prime}27^{\prime \prime}.60$, error radius (90\%) = 4$^{\prime}$.95;
(99\%) = 7$^{\prime}$.72)}}\\
   &   &   &   &   &   &   \\
\#1 (in 90\%) & $18^{\rm h}07^{\rm m}53^{\rm s}.18$ & $+$$38^\circ22^{\prime}41^{\prime \prime}.91$ & 6.12 &
$5.33\pm1.60$ &  --   &  1/4207\\
\hline
\multicolumn{7}{c}{\textbf{XMMSL1 J182831.8$-$022901 (R.A.(J2000) = $18^{\rm h}28^{\rm m}26^{\rm s}.16$,
Dec(J2000) = $-$$02^\circ29^{\prime}14^{\prime \prime}.12$, error radius (90\%) = 3$^{\prime}$.74;
(99\%) = 5$^{\prime}$.83)}}\\
    &   &   &   &   &   &   \\
\#1 (in 90\%) & $18^{\rm h}28^{\rm m}31^{\rm s}.09$ & $-$$02^\circ29^{\prime}06^{\prime \prime}.66$ & 3.89 &  
$45.12\pm3.90$ &  $26.68\pm3.00$   & 7/5266  \\
\hline
\multicolumn{7}{c}{\textbf{SWIFT J1839.1$-$5717 (R.A.(J2000) = $18^{\rm h}39^{\rm m}03^{\rm s}.36$, Dec.(J2000) =
 $-$$57^\circ14^{\prime}56^{\prime \prime}.40$, error radius (90\%) = 4$^{\prime}$.69;
(99\%) = 7$^{\prime}$.32)}}\\
   &   &   &   &   &   &   \\
\#1 (in 90\%) & $18^{\rm h}39^{\rm m}06^{\rm s}.37$ & $-$$57^\circ15^{\prime}05^{\prime \prime}.83$ & 3.57 &
$113.60\pm3.50$ &  $61.75\pm2.60$  &  2/16039\\
\#2 (border of 90\%) & $18^{\rm h}38^{\rm m}41^{\rm s}.94$ & $-$$57^\circ18^{\prime}37^{\prime \prime}.78$ & 3.89 &
$5.95\pm0.89$ &  --   &   \\
\hline
\multicolumn{7}{c}{\textbf{IGR J20310$+$3835 (R.A.(J2000) = $20^{\rm h}31^{\rm m}01^{\rm s}.20$, Dec.(J2000) =
$+$$38^\circ34^{\prime}33^{\prime \prime}.60$, error radius (90\%) = 4$^{\prime}$.54;
(99\%) = 7$^{\prime}$.08)}}\\
    &   &   &   &   &   &   \\
\#1 (in 90\%) & $20^{\rm h}30^{\rm m}55^{\rm s}.27$ & $+$$38^\circ33^{\prime}44^{\prime \prime}.14$ & 4.35 &
$19.96\pm2.70$ &   $13.61\pm2.20$   &  7/4827  \\
\#2 (in 99\%) & $20^{\rm h}30^{\rm m}45^{\rm s}.00$ & $+$$38^\circ39^{\prime}07^{\prime \prime}.46$ & 6.21 &
$5.50\pm1.60$ &  --    & \\
\hline
\multicolumn{7}{c}{\textbf{1SWXRT J230642.8$+$550817 (R.A.(J2000) = $23^{\rm h}06^{\rm m}54^{\rm s}.48$, Dec(J2000) =
$+$$55^\circ10^{\prime}22^{\prime \prime}.80$, error radius (90\%) = 3$^{\prime}$.79;
(99\%) = 5$^{\prime}$.91)}}\\
   &   &   &   &   &   &   \\
\#1 (in 90\%) & $23^{\rm h}06^{\rm m}42^{\rm s}.33$ & $+$$55^\circ08^{\prime}18^{\prime \prime}.91$ & 4.51 & 
$96.50\pm13.00$ &  $35.32\pm8.10$  &    2/946 \\
\hline
\hline
\end{tabular}
\end{center}
\end{table*}

\section{Results}
To investigate the nature of each potential counterpart, we browsed various on-line archives 
such as NED (NASA/IPAC Extragalactic Database), HEASARC (High Energy Astrophysics Science 
Archive Research Center) and SIMBAD (Set of Identifications, Measurements, and Bibliography 
for Astronomical Data) in search of radio, infrared, optical, and UV counterparts within the 
XRT positional uncertainty. When relevant, we also discuss the association with objects 
reported in the \emph{ROSAT} and \emph{XMM-Newton} Slew
catalogues (Voges et al.1999; 
Saxton et al. 2008)\footnote{More information for these two catalogues are available at:\\ 
\url{https://heasarc.gsfc.nasa.gov/W3Browse/rosat/rassbsc.html};   
\url{https://heasarc.gsfc.nasa.gov/W3Browse/xmm-newton/xmmslewcln.html.}}

In Table~\ref{tab2} we list the 14 IBIS sources analysed here together with their coordinates 
and relative uncertainty (90\% c.l.) as listed in Bird et al. (2016). For each of these 
gamma-ray emitters, we then report the coordinates and relative uncertainties (at 90\% c.l.) 
of all sources detected by XRT within the 90\% and 99\% IBIS positional uncertainties, the 
count rate in both the 0.3--10 and 3--10 keV energy range, the number of X-ray observations 
analysed and the total on-source exposure time. For each IBIS source, in Table~\ref{tab3} we report 
those XRT objects for which an optical and IR counterpart was found; references for the 
databases used are reported at the end of Table~\ref{tab3}.

The results of the spectral analysis are shown in Table~\ref{tab4} where we report the 
Galactic column density in the source direction (Kalberla et al. 2005) and the best-fit 
parameters (intrinsic column density, power law photon index, $\chi^{2}$ or C-stat versus 
degrees of freedom and 2--10 keV flux). Spectra and data-to-model ratio are shown for only 
those sources treated with the $\chi^{2}$ statistic.

In the following, we discuss each \emph{INTEGRAL}/IBIS source (as reported in Bird et al. 
2016) and briefly analyse the overall properties found for each candidate counterpart.

\subsection{\bfseries{XMMSL1 J030715.5$-$545536}\\ 
(\textit{Detected as a persistent source})}

This is the only IBIS source for which XRT follow-up observations do not reveal the presence 
of X-ray sources in the region surrounding the high-energy emitter. However, within the 90\% 
IBIS positional uncertainty there is an \emph{XMM-Newton} Slew source (XMMSL1 
J030715.5$-$545536) that is detected at around 2$\sigma$ c.l. in the 0.2--12 keV energy band 
with a flux of $1.5\times10^{-12}$ erg cm$^{-2}$ s$^{-1}$. The XRT upper limit in the same 
energy range is $\sim$$1.2\times10^{-14}$ erg cm$^{-2}$ s$^{-1}$, which indicates significant 
flux variability over the period covered by the \emph{XMM-Newton} Slew (November 2010) and XRT 
(March 2016) pointings. Within the restricted positional uncertainty of the \emph{XMM-Newton} 
Slew detection (5.1 arcsec), we do not find any optical or IR counterpart. We note that 4$\%$ 
of the sources in the clean Slew Survey catalogue are expected to be spurious from statistical 
considerations (see Saxton et al. 2008), which suggests that this possibility cannot be 
totally discounted. Alternatevely, an X-ray source is present inside the IBIS positional 
uncertainty but variable over time, more strongly at lower than at higher energies given the 
persistent nature of the source within the \emph{INTEGRAL} database used by Bird et al. 
(2016). If so, and considering the source high Galactic latitude ($b= -53^{\circ}$), this 
X-ray detection could be an AGN maybe of the blazar type. Only X-ray monitoring of the source 
can provide some clues on its nature, while spectroscopy of the only possible X-ray 
counterpart can confirm or not its association with the gamma-ray source.

\subsection{\bfseries{SWIFT J0800.7$-$4309}\\
(\textit{Detected as a persistent source})} 

\begin{figure}
\includegraphics[width=1.0\linewidth]{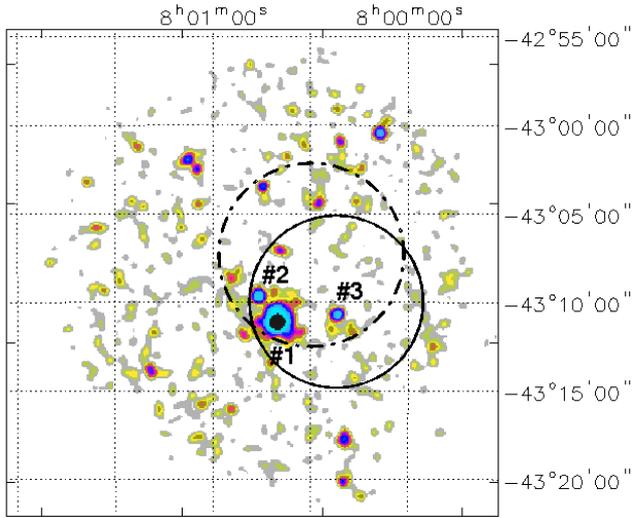}
\caption{XRT 0.3--10 keV image of the region surrounding SWIFT J0800.7$-$4309.
Three X-ray sources are detected within the 90\% IBIS and BAT positional uncertainties (black circle
and black-dashed-dotted circle, respectively).}
\label{fig1}
\end{figure}

This source is also unclassified in the 70-month \emph{Swift}/BAT survey (Baumgartner et al. 
2013). The observations performed with XRT show the presence of three X-ray sources within the 
90\% IBIS/BAT positional uncertainty (see Table~\ref{tab2} and Figure~\ref{fig1}), two of 
which (source \#1 and \#2) are still detected above 3 keV.

\begin{figure}
\includegraphics[width=1.1\linewidth]{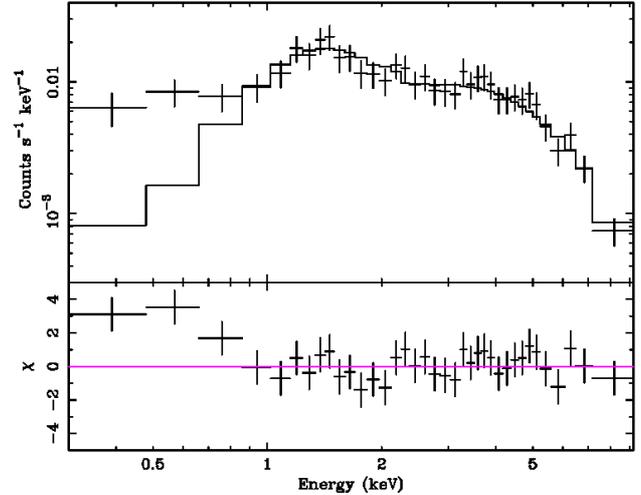}
\caption{XRT spectrum of SWIFT J0800.7$-$4309 source \# 1 fitted with our basic model 
(\emph{upper panel}); residuals to this
model are in units of $\sigma$ (\emph{lower panel}).}
\label{fig2}
\end{figure}

In more detail, for source \#1, which is the brightest object even above 3 keV (15.2$\sigma$ 
c.l.), we found a single optical and IR counterpart within its error circle (see 
Table~\ref{tab3}). Recent optical follow-up observations (Rojas et al. 2016) indicate that 
this is a Cataclysmic Variable binary system source (CV). Our basic model ($\Gamma \sim 0.8$ 
and 2--10 keV flux of $\sim$$5\times10^{-12}$ erg cm$^{-2}$ s$^{-1}$, see Table~\ref{tab4}) 
does not yield a good fit to the XRT data ($\chi^{2}/d.o.f. = 43.3/35$), as an excess below 1 
keV is clearly visible in the data-to-model ratio (see Figure~\ref{fig2}). This feature, which 
is expected to be observed in CVs (see Landi et al. 2009 and references therein), can be 
modelled with a blackbody component. Unfortunately, the statistical quality of the X-ray data 
does not allow us to place any constraint on this component.

A single optical and IR counterpart (see Table~\ref{tab3}) was also found within the 
positional uncertainty of source \#2, which is still detected above 3 keV but only at 
3.4$\sigma$ c.l.. For this object, the XRT spectral analysis yields a 2--10 keV flux 
of $\sim$$2\times10^{-13}$ erg cm$^{-2}$ s$^{-1}$ and a photon index around 1.2 (see 
Table~\ref{tab4}).

The X-ray brightness of source \#1 and its optical classification argue in favour of its 
association with SWIFT J0800.7$-$4309, but optical spectroscopy of source \#2 is required 
before its contribution to the soft gamma-ray emission can totally be disregarded.

\subsection{\bfseries{SWIFT J0924.2$-$3141}\\
(\textit{Detected as a persistent source})}

\begin{figure}
\includegraphics[width=1.\linewidth,angle=0]{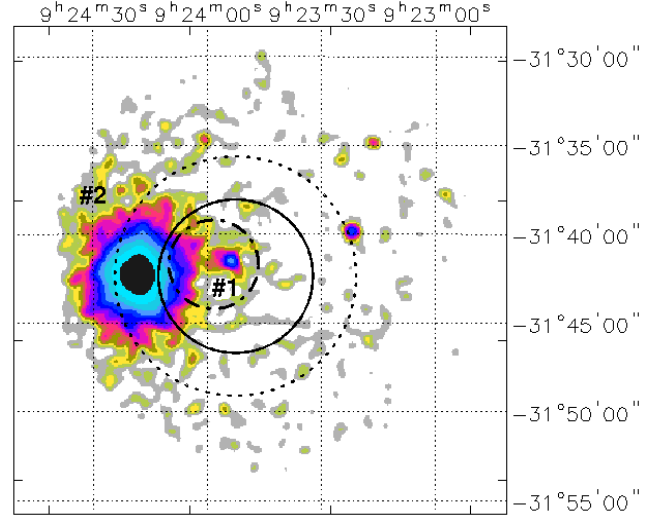}
\caption{0.3--10 keV XRT image of the SWIFT J0924.2$-$3141 field. 
Two sources are detected by XRT: source \#1 and \#2 lie within the 90\% (black circle) and 
99\% (black-dotted circle) positional uncertainties, respectively. 
While source \#1 is contained within the 90\% BAT error circle (black-dashed-dotted circle), 
source \#2 lies just outside it.}
\label{fig3}
\end{figure}

Also this source is reported in the \emph{Swift}/BAT 70-month survey (Baumgartner et al. 
2013). XRT follow-up observations reveal the presence of two objects (see Figure~\ref{fig3} 
and Table~\ref{tab2}) in the region surrounding this high-energy emitter. Their positions are 
compatible with either the 90\% (\# 1) or the 99\% (\#2) IBIS positional uncertainties, but 
only source \#1 is located within the 90\% BAT error circle.

Source \# 1 is detected at 5.7$\sigma$ c.l. both in 0.3--10 keV energy range and above 3 keV. 
The source is visible at 2.4$\sigma$ c.l. up to 6.7 keV. Its optical/IR counterparts are 
listed in Table~\ref{tab3}: both are associated with the galaxy 2MASX J09235371$-$3141305, 
classified as a Seyfert 1.8 at $z=0.042$ in the Veron \& Veron (13$^{th}$ edition) catalogue 
(2010). The XRT localization is also compatible with a NVSS radio source belonging to the NRAO 
VLA Sky Survey (NVSS; Condon et al. 1998), namely NVSS J092353$-$314126, listed with a 20 cm 
flux density of $4.4\pm0.5$ mJy.  This source was proposed by Baumgartner et al. (2013) as the 
counterpart to SWIFT J0924.2$-$3141. Furthermore, Ricci et al. (2015), combining XRT and BAT 
spectra, have recently suggested that this may be a Compton thick AGN. The XRT data, although 
of poor statistical quality, require a double power-law model, with the primary component 
absorbed by an intrinsic column density and the secondary component, having the same photon 
index (frozen to 1.8) of the primary one, passing only through the Galactic absorption. The 
intrinsic $N_{\rm H}$, albeit poorly constrained, is found to be around $8\times 10^{23}$ 
cm$^{-2}$ and compatible with the Compton thick regime within uncertainties, while the 
2--10 keV flux is $\sim$$1.4\times10^{-12}$ erg cm$^{-2}$ s$^{-1}$ (see Table~\ref{tab4}).

\begin{figure}
\includegraphics[width=1.1\linewidth]{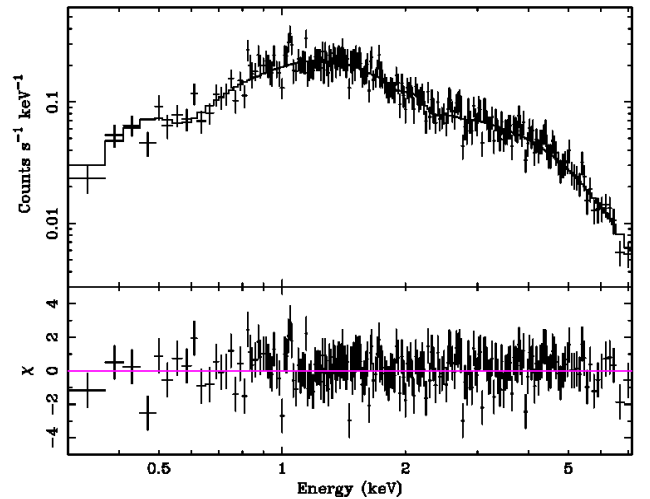}
\caption{XRT spectrum of SWIFT J0924.2$-$3241 source \# 2 fitted with our basic model plus a 
black body component 
(\emph{upper panel}); residuals to this model are in units of $\sigma$ (\emph{lower panel}).}
\label{fig4}
\end{figure}

Source \#2, which lies just outside the 90\% IBIS error circle but within the 99\% one, is 
much brighter than object \#1, as it is detected at around 126$\sigma$ and 72$\sigma$ c.l. in 
the 0.3--10 keV energy band and above 3 keV, respectively. Its XRT position is compatible with 
an USNO$-$A2.0 object; no IR counterpart has been found (see Table~\ref{tab3}). Browsing the 
HEASARC archive, we find that this source was detected by various X-ray instruments like 
\emph{Chandra} as CXO J092418.2$-$314217, \emph{Beppo}SAX Wide Field Camera as SAXWFC 
J0924.3$-$3142.4, and \emph{ROSAT} as the bright source 1RXS J092418.0$-$314212. It also 
coincides with an \emph{XMM-Newton} Slew source (XMMSL1 J092418.4$-$314219), which is detected 
at 14.3$\sigma$ c.l., with a 0.2--12 keV flux of $\sim$$1.42\times10^{-10}$ erg cm$^{-2}$ 
s$^{-1}$. Historically, the source may have also been observed by \emph{ARIEL V} and 
\emph{UHURU} with a 2--6 keV flux around $8 \times10^{-11}$ erg cm$^{-2}$ s$^{-1}$ and 
reported in the \emph{HEAO--1} A3 MC LASS catalogue (H0922$-$374), where it is classified as 
an X-ray binary in the Galaxy, due to the fact that the X-ray flux is by far too bright for an 
AGN of magnitude $V \sim21$. Furthermore, the X-ray/optical fluxes and ASM colours most likely 
resemble low luminosity ultra compact binaries. By fitting the average XRT spectrum with a 
simple power law, we found a photon index $\Gamma \sim 1.4$ and a 2--10 keV flux of 
$\sim$$2.4\times10^{-10}$ erg cm$^{-2}$ s$^{-1}$. The addition of a blackbody component 
provides a significant improvement of the fit ($\Delta\chi^{2} = 40.0$ for two d.o.f. less), 
yielding a temperature $kT \sim 1.1$ keV, a photon index of $\sim 1.7$ and a 2--10 keV flux of 
$\sim$$1.9\times10^{-10}$ erg cm$^{-2}$ s$^{-1}$ (see Figure~\ref{fig4}). Analysing single XRT 
observations we find variability by a factor of 1.7 over a year time-scale and by a factor of 
2 in comparison with the \emph{XMM-Newton} Slew flux measurement.

\begin{figure*}
\includegraphics[width=0.45\linewidth,angle=0]{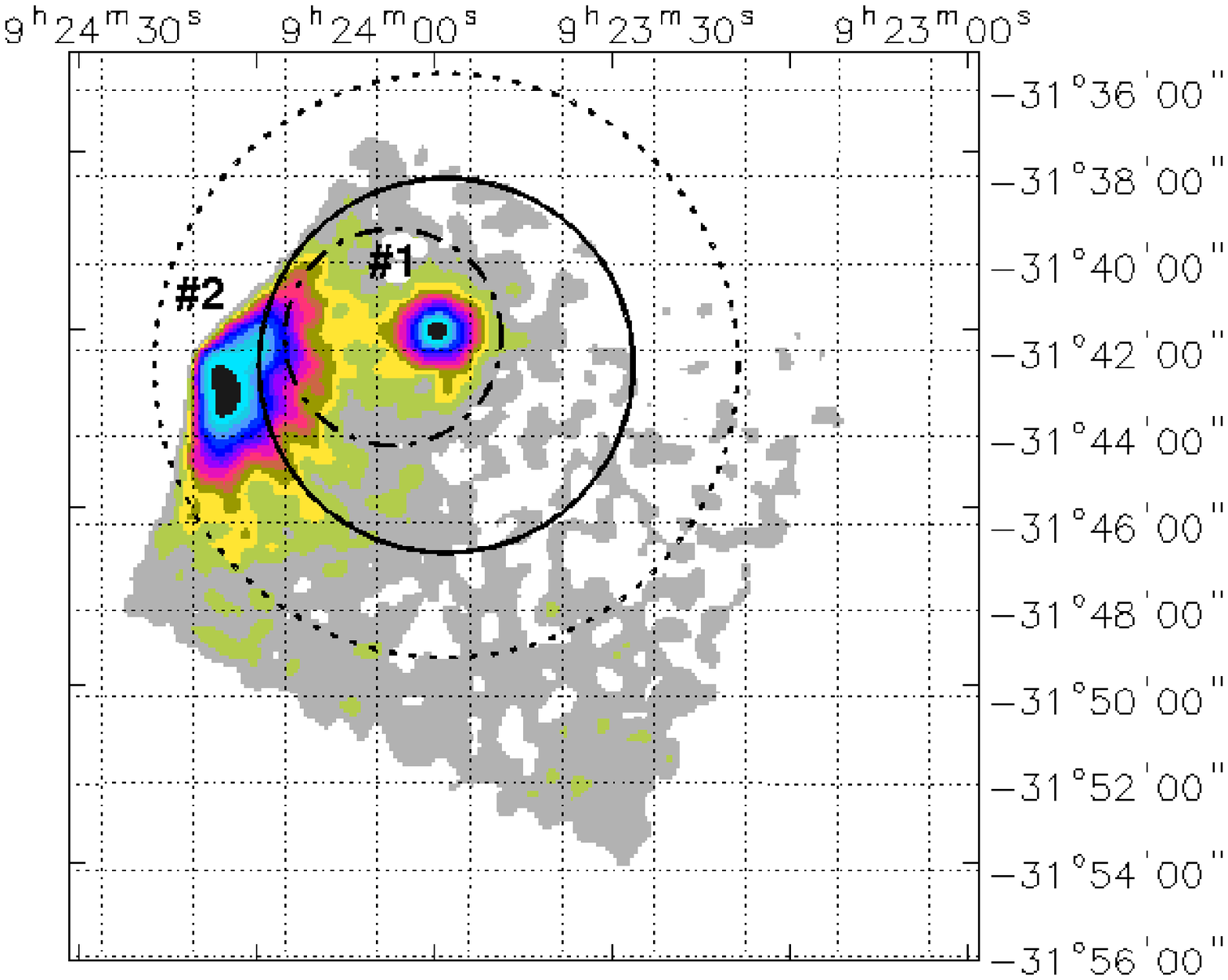}
\includegraphics[width=0.45\linewidth,angle=0]{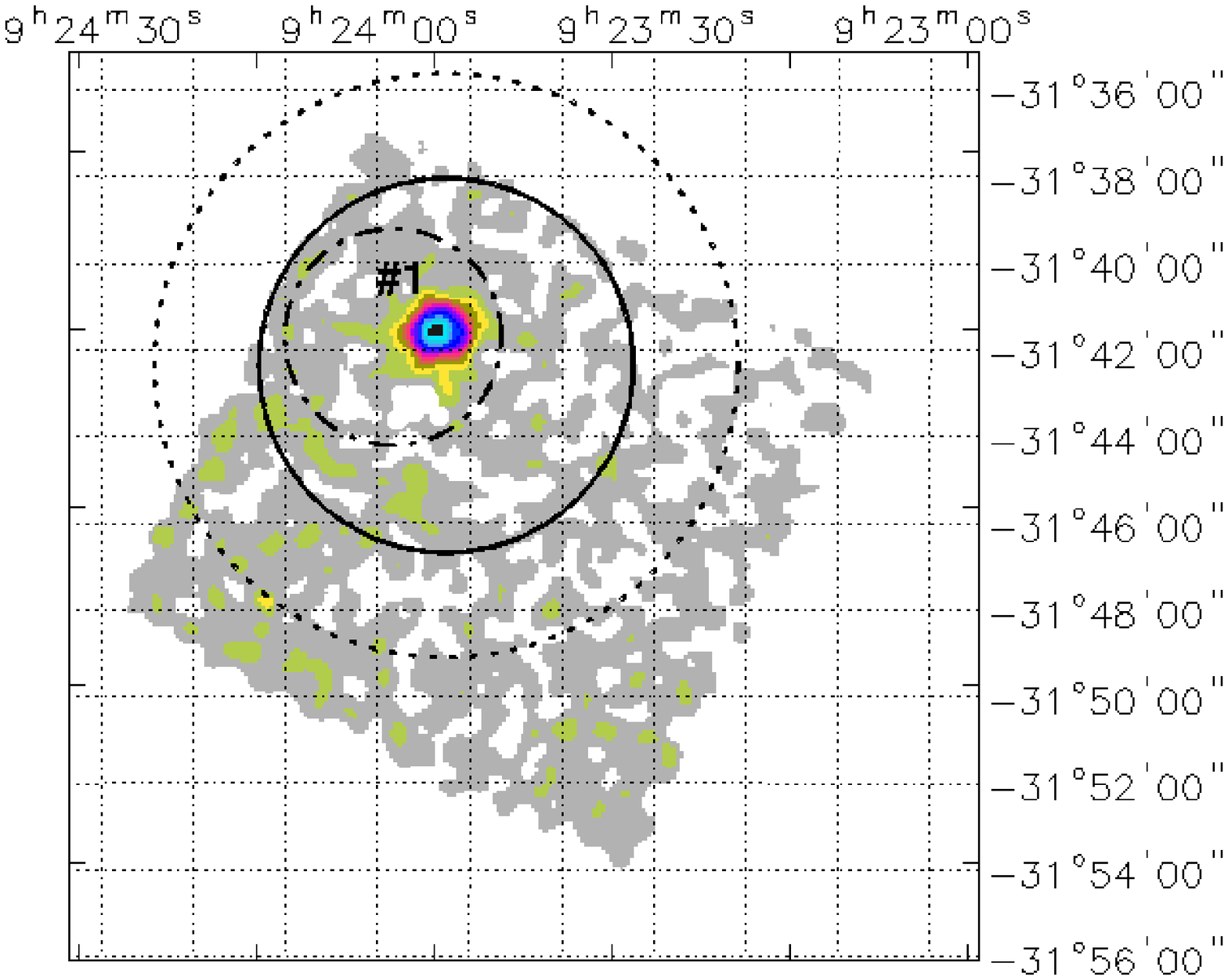}
\caption{NuSTAR images of the SWIFT J0924.2$-$3141 field. 
\emph{Left panel}: image in the 3--79 keV energy
band, where the two sources detected by XRT are clearly visible. 
\emph{Right panel}: image in the 15--79 keV 
energy band, where only source \#1 is still detected by NuSTAR.} 
\label{fig5}
\end{figure*}

As can be seen in Figure~\ref{fig3}, the XRT data are not able alone to discriminate between 
source \#1, located within the 90\% IBIS/BAT error circle, very hard but less bright and 
source \#2, 100 times brighter but located outside the 90 \% IBIS positional uncertainty. 
Luckily, NuSTAR has recently performed an observation of this sky region: although both 
sources are clearly visible in the 3--79 keV image, only source \#1 is present above 15 keV 
(see Figure~\ref{fig5})\footnote{NuSTAR images were extracted starting from the event files 
available in the ASI/ASDC data archive available at:\\ 
\url{http://www.asdc.asi.it/mmia/index.php?mission=numaster}.} and is therefore the real 
counterpart of this IBIS/BAT detection. Source \#2 remains, however, an interesting object to 
study and one for which optical follow-up observations would be very useful. Fitting 
the NuSTAR spectrum of source \#1\footnote{For this spectral extraction, we follow the 
prescription used by Malizia et al. (2016) for other AGN observed by NuSTAR.} with an absorbed 
power law provides a good fit ($\chi^{2}/d.o.f. = 58.76/56$) to the data and the following 
best-fit parameters: a photon index of $1.5\pm0.2$, a column density of $(5.9 
\pm1.7)$$\times10^{23}$ cm$^{-2}$ and a 2--10 keV flux of $\sim$$1.3\times10^{-12}$ erg 
cm$^{-2}$ s$^{-1}$; this indicates that source \#1 is heavily absorbed, but not quite Compton 
thick.

\subsection{\bfseries{IGR J14059$-$6116}\\
(\textit{Detected in a 2198.6-day outburst from MJD = 52980.4})} 
 
\begin{figure}
\includegraphics[width=1.0\linewidth]{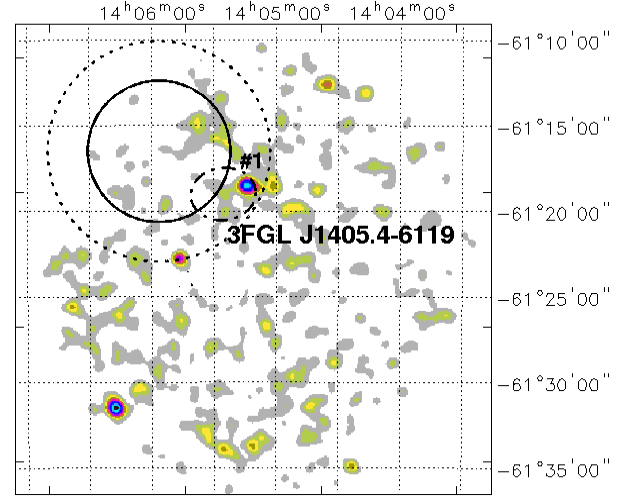}
\caption{XRT 0.3--10 keV image of the region surrounding IGR J14059$-$6116.
The only XRT detection (source \#1) is located inside the 99\% IBIS error circle 
(black-dotted circle) 
and also lies within the error ellipse of the \emph{Fermi} source 3FGL J1405.4$-$6119 
(black-dashed-dotted ellipse).}
\label{fig6}
\end{figure}

According to the bursticity method (see Bird et al. 2016 for details), this source was 
detected during activity in a set of data corresponding to roughly 2200 days over the period 
December 7, 2003 to December 14, 2009 (MJDs 52980.45 -- 55179.04). XRT follow-up observations 
were instead carried out at a later time, more specifically during 2011 (October 4, 8 and 
November 1), 2012 (September 21) and 2015 (February 19 and April 27). It is therefore possible 
that the X-ray pointings missed the active phase seen by IBIS above 20 keV and observed the 
source in a more typical X-ray state. Despite this note of caution, the combined XRT image of 
this sky region (Figure~\ref{fig6}) shows the presence of only one source whose position is 
compatible with the 99\% IBIS positional uncertainty (see Table~\ref{tab2}).

This source, which is detected around 4$\sigma$ and 3.4$\sigma$ c.l. in the range 0.3--10 and 
3--10 keV, respectively, is also included in the positional uncertainty of the source 3FGL 
J1405.4$-$6119 (black-dashed-dotted ellipse in Figure~\ref{fig6}) 
belonging to the third \emph{Fermi} 
Large Area Telescope catalogue (Acero et al. 2015). The XRT positional uncertainty contains a 
single infrared (2MASS/WISE) counterpart that is not reported in any optical catalogue (see 
Table~\ref{tab3}).

From the XRT data, we can only infer a 2--10 keV flux roughly around $3\times10^{-13}$ 
erg cm$^{-2}$ s$^{-1}$, by assuming our basic model with the photon index frozen to 1.8. The 
source has also been observed by \emph{Chandra} on September 19, 2013: it is listed in the 
\emph{Chandra} ACIS GSG Point-Like X-Ray Source Catalog (Wang et al. 20016) as CXOGSG 
J140514.4$-$611827 (only 1 arcsec uncertainty) with a 0.3--8.0 keV flux of $2.6\times10^{-13}$ 
erg cm$^{-2}$, i.e. similar to the XRT one.

The detection of the XRT source inside the 3FGL error ellipse and its connection with the IBIS 
object is particularly interesting and worth investigating. Unfortunately, at this stage, the 
limited multi-waveband information prevents us from finding any secure clues on the nature of 
the source. The lack of a radio emission within the XRT potional uncertainty together 
with the allWISE colours ($W1-W2 = 0.381$, and $W2-W3 = 1.281$) of the IR counterpart, 
indicates that a blazar interpretation is unlikely (Masetti et al. 2013). Moreover, the 
source is located at low Galactic latitude ($b= +0.29^{\circ}$), i.e. on the Galactic plane, 
and 3FGL J1405.4$-$6119 has already been reported as a pulsar candidate by Lee et al. (2012) 
and more recently by Saz-Parkinson et al. (2016).
 
Clearly, this remains an object of uncertain type, but its association with GeV emission is of 
particular interest and future dedicated high-energy observations and optical spectroscopy of 
the XRT source may shed light on its ultimate nature.

\subsection{\bfseries{1RXS J145959.4$+$120124}\\
(\textit{Detected in a 48.7-day outburst from MJD = 54640.7})}

\begin{figure}
\includegraphics[width=1.0\linewidth]{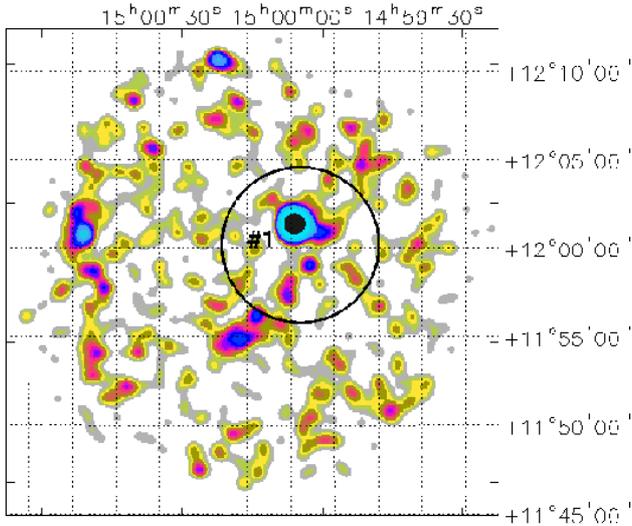}
\caption{XRT 0.3--10 keV image of the region surrounding 1RXS J145959.4$+$120124.
Only one X-ray source is detected within the 90\% IBIS positional uncertainty (black circle).}
\label{fig7}
\end{figure}

This is another IBIS source detected with the bursticity method; it was detected during a 
49-day outburst starting from June 23, 2008. The XRT observations were instead performed on 
March (15 to 18) 2016. As shown in Figure~\ref{fig7}, there is only one X-ray source within 
the IBIS 90\% positional uncertainty; it is detected around 12$\sigma$ and 5$\sigma$ in the 
0.3--10 and 3--10 keV energy range respectively. It has a counterpart in an \emph{XMM-Newton} 
Slew object (XMMSL1 J145959.6$+$120131, $6.^{\prime \prime}$2 error radius) that is also 
reported in the \emph{ROSAT} Bright source catalogue as 1RXS J145959.4$+$120124 ($11^{\prime 
\prime}$ error radius). By \emph{XMM-Newton} it is detected at 3.7$\sigma$ c.l. in the 0.2--12 
keV energy range with a flux of 5.32$\times10^{-12}$ erg cm$^{-2}$ s$^{-1}$, 60$\%$ of which 
is above 2 keV\footnote{\emph{XMM-newton} Slew observation date is February 6, 2002.}. In the 
\emph{XMM-Newton} Slew catalogue, the source is associated with HD 132658/TYCHO 
922$-$865$-$1\footnote{See more information at:\\ 
\url{http://www.astrostudio.org/xhip.php?hip=73397}.}, a bright star of spectral type F5 D, 
whose location is also compatible with the positional uncertainty of the \emph{ROSAT} 
detection.

\begin{figure}
\includegraphics[width=1.1\linewidth]{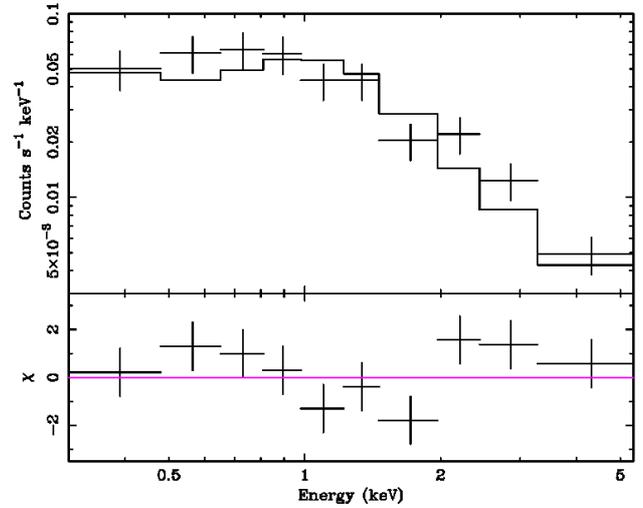}
\caption{XRT spectrum of 1RXS J145959.4$+$120124 fitted with our basic model  
(\emph{upper panel}); residuals to this model are in units of $\sigma$ (\emph{lower panel}).}
\label{fig8}
\end{figure}

The spectral analysis of the average XRT spectrum provides a photon index of $\sim 1.9$ and a 
2--10 (0.2--12 keV) keV flux of 2.4 (5.0) $\times10^{-12}$ erg cm$^{-2}$ s$^{-1}$ (see 
Figure~\ref{fig8} and Table~\ref{tab4}); the XRT flux ranges from 1.2 to 2.4$\times10^{-12}$ 
erg cm$^{-2}$ s$^{-1}$. Therefore, spectral analysis of each single observation indicates 
variability by a factor of $\sim2$ on a four-day time-scale, while comparison of the average 
XRT flux with the \emph{XMM-Newton} Slew one indicates similar flux levels.

\begin{figure}
\includegraphics[width=1.\linewidth,angle=0]{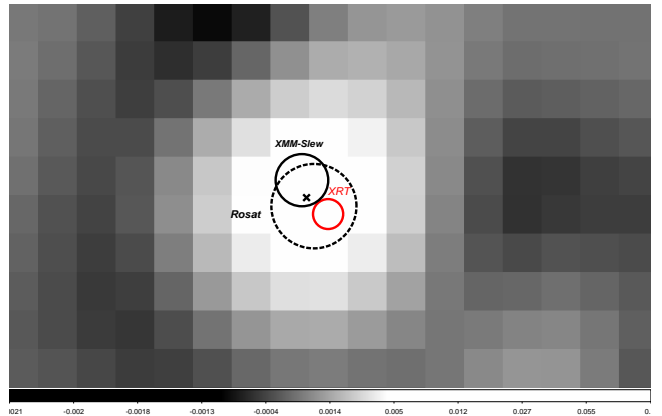}
\caption{NVSS image of the 1RXS J145959.4$+$120124 field, where we plot the position of the star
HD 132658/TYCHO 922$-$865$-$1 (black cross), 
as well as the XRT, \emph{ROSAT}, and \emph{XMM-Newton} Slew  
positional uncertainties (red, black and black-dotted circles, respectively).}
\label{fig9}
\end{figure}

The X-ray source may also be associated with radio emission since a NVSS and FIRST (Faint 
Images of the Radio Sky at 20 cm, Helfand et al. 2015) source is listed nearby. The source, 
named NVSS J145959$+$120126/FIRST J145959.3$+$120125, has a 20 cm flux density of 6.5$\pm$0.4 
and 8.88$\pm$0.15 mJy in the two catalogues, respectively. The NVSS image of this sky region 
is shown in Figure~\ref{fig9} where we plot the XRT, \emph{ROSAT}, and \emph{XMM-Newton} Slew 
error circles, as well as the position of the star HD 132658/TYCHO 922$-$865$-$1. 
Figure~\ref{fig9} emphasises that this is clearly a difficult case: on the one hand the star 
is formally outside the XRT positional uncertainty, but it is compatible with the \emph{ROSAT} 
and \emph{XMM-Newton} Slew error circles and it is spatially coincident with the radio source; 
on the other hand it is difficult to explain radio and soft gamma-ray emission from an F type 
class object. Alternatevely, one must consider the possibility that the star is a chance 
association and that the true counterpart is a background object masked by the brightness of 
the star. In this case, the X-ray properties, the location of the source at high Galactic 
latitudes and the presence of radio emission suggest an extragalactic nature, i.e. a variable 
AGN behind HD 132658/TYCHO 922$-$865$-$1. Unfortunately, optical/IR follow-up observations 
cannot help in this case, since the bright star prevents the detection of objects nearby or 
behind it, thus making this case difficult to solve.

\subsection{\bfseries{IGR J15038$-$6021}\\
(\textit{Detected  as a persistent source})}

\begin{figure}
\includegraphics[width=1.0\linewidth]{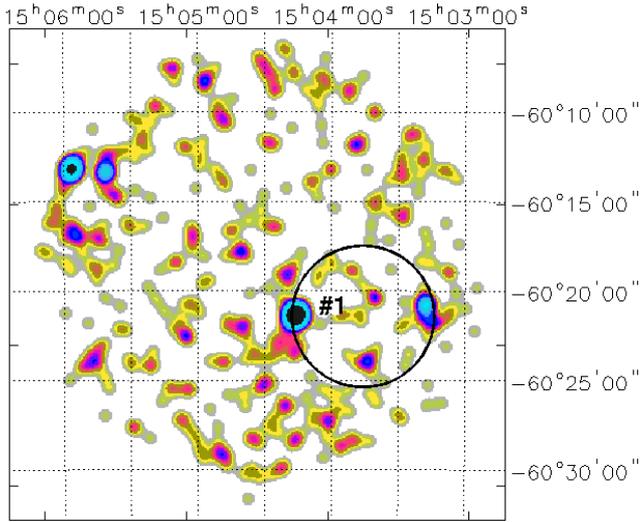}
\caption{XRT 0.3--10 keV image of the region surrounding IGR J15038$-$6021.
Only one source is detected by XRT within the 90\% IBIS error circle (black circle).}
\label{fig10}
\end{figure}

In this case, only one X-ray source is clearly visible inside the 90\% IBIS error circle as 
listed in Table~\ref{tab2} and depicted in Figure~\ref{fig10}. It is detected at 5$\sigma$ and 
3.4$\sigma$ c.l., in the 0.3--10 and 3--10 keV energy range, respectively. Because of 
the poor quality of the XRT data, we can only infer a 2--10 keV flux of 
$\sim$$1.6\times10^{-12}$ erg cm$^{-2}$ s$^{-1}$, by freezing the photon index to 1.8.

Two optical and one IR counterparts were found within the XRT positional uncertainty (see 
Table~\ref{tab3}).

The hardness of this source in X-rays and the lack of other X-ray detections suggest that it 
is a likely association with IGR J15038$-$6021. Spectroscopy of both optical candidates are 
therefore encouraged to pinpoint which of the two is the true counterpart and assess its 
nature.

\subsection{\bfseries{IGR J16447$-$5138}\\
(\textit{Detected in a 35.7-day outburst from MJD = 54141.5})}

This source was detected during an outburst lasting roughly 5 weeks starting from February 10, 
2007; the XRT pointings were made at a much later time (March 4--6, 2016) and may be related 
to a more inactive state of the source.
 
\begin{figure}
\includegraphics[width=1.0\linewidth]{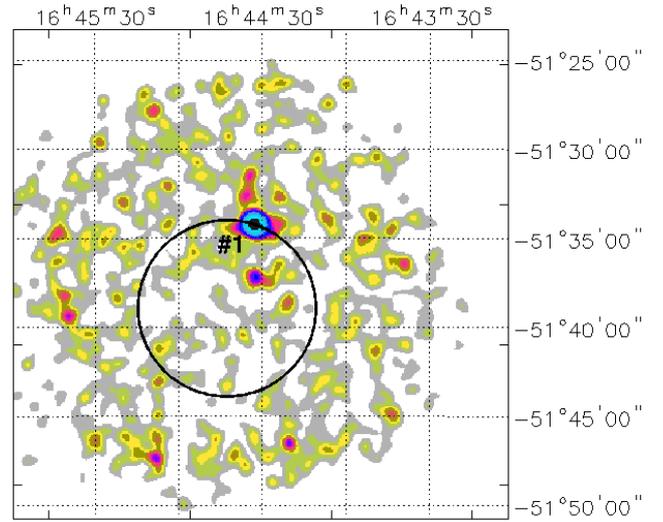}
\caption{XRT 0.3--10 keV image of the region surrounding IGR J16447$-$5138.
The only XRT detection lies at the border of the
90\% IBIS positional uncertainty (balck error circle).}
\label{fig11}
\end{figure}

The only X-ray object seen in the region surrounding the IBIS emitter is located at the border 
of the 90\% IBIS error circle (see Figure~\ref{fig11}). It is detected at 13.4$\sigma$ and 
7.1$\sigma$ c.l. in the 0.3--10 and 3--10 keV energy range. The source is also reported as an 
\emph{XMM-Newton} Slew object XMMSL1 J164433.3$-$513420 with a 0.2--12 keV flux of 
1.30$\times10^{-12}$ erg cm$^{-2}$ s$^{-1}$.

\begin{figure}
\includegraphics[width=1.1\linewidth]{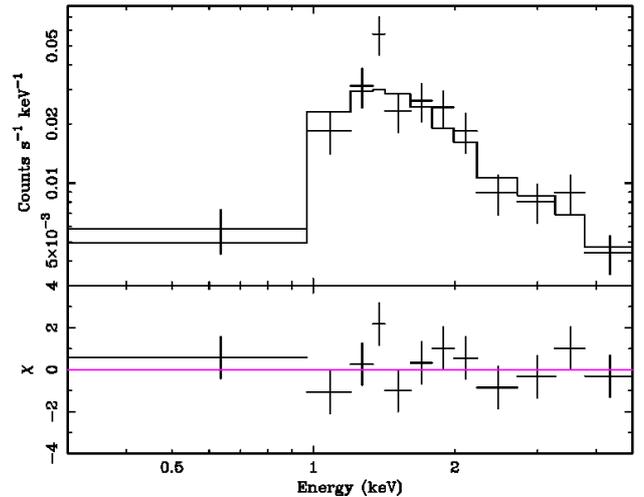}
\caption{XRT spectrum of IGR J16447$-$5138 fitted with our basic model plus intrinsic absorption 
(\emph{upper panel}); residuals to this model are in units of $\sigma$ (\emph{lower panel}).}
\label{fig12}
\end{figure}

The XRT spectrum is well modelled with an absorbed power law ($N_{\rm{H(int)}} < 0.6 
\times10^{22}$ cm$^{-2}$), yielding a photon index $\Gamma \sim 2.0$ and 2--10 keV flux of 
$\sim$$3.9\times10^{-12}$ erg cm$^{-2}$ s$^{-1}$ (see Figure~\ref{fig12} and 
Table~\ref{tab4}). The source shows flux variability by a factor of 1.5 over a two-day 
time-scale comparing closeby \emph{Swift} pointings. The XRT flux is also higher by a factor 
of 3.5 than the \emph{XMM-Newton} Slew one in the same energy range. In Table~\ref{tab3} we 
report the only optical and IR counterpart to the XRT source.

We note that this object is on the Galactic plane and has been suggested to be a Young Stellar 
Object (YSO) by Marton et al. (2016) on the basis of its WISE and 2MASS photometric data. 
above 20 keV during flaring events, while 
coronal activity. However, we note that extragalactic sources, especially galaxies with 
ongoing star formation or active galactic nuclei, show similar infrared spectral shape to that 
of YSOs and may be variable over time; this leaves the final answer on which is the nature of 
this source to optical follow-up spectroscopic observations and eventually to X-ray monitoring 
campaigns.

\subsection{\bfseries{IGR J17508$-$3219}\\
(\textit{Detected as a persistent source})}

\begin{figure}
\includegraphics[width=1.0\linewidth]{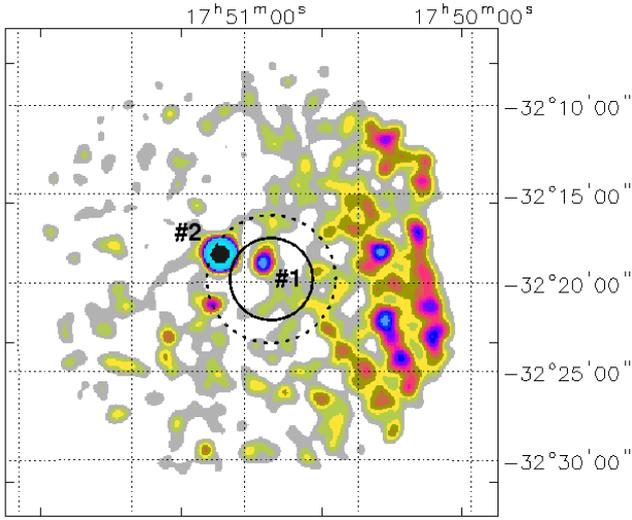}
\caption{XRT 0.3--10 keV image of the region surrounding IGR J17508$-$3219.
Two sources are detected by XRT: source \#1
and \#2 lie within the 90\% and 99\% IBIS positional uncertainty 
(black and black-dotted circles), respectively.}
\label{fig13}
\end{figure}

In this case, two X-ray detections are revealed by XRT in the
region surrounding IGR J17508$-$3219 as can be seen in Figure~\ref{fig13} 

Source \#1, which is located within the 90\% IBIS error circle, is the weakest of the two, but 
also the only one detected above 3 keV (at 3.1$\sigma$ c.l., see Table~\ref{tab2}). In 
Table~\ref{tab3} it has a single infrared counterpart (GLIMPSE G357.6922$-$02.7203), 
which is not seen at optical wavelengths; because of the low statistical quality of the X-ray 
data the spectral parameters are poorly constrained. The fit with our basic model 
yields a flat photon index around 0.6 and a 2--10 keV flux around 8$\times10^{-13}$ erg 
cm$^{-2}$ s$^{-1}$; fixing the photon index to 1.8 provides an intrinsic absorption around 
$1\times10^{22}$ cm$^{-2}$.

Source \#2, which lies within the 99\% IBIS positional uncertainty, is detected at 
13.4$\sigma$ c.l. in the 0.3--10 keV energy range, but it is not revealed above 3 keV. This 
source coincides with the \emph{XMM-Newton} Slew source (XMMSL1 J175106.4$-$321824, error 
radius 2 arcsec), detected at 4.4$\sigma$ c.l. with a 0.2--12 keV flux of 7.48$\times10^{-12}$ 
erg cm$^{-2}$ s$^{-1}$. It has also been associated with the \emph{ROSAT} Bright object 1RXS 
J175106.2$-$321836 (error radius 10 arcsec), itself identified as the star HIP 87368 (HD 
162186) of spectral type G3IV(e). 

The XRT data are best modelled with a bremsstrahlung component having $kT = 
0.31^{+0.11}_{-0.13}$ keV and a 2--10 keV flux of around $8\times10^{-14}$ erg cm$^{-2}$ 
s$^{-1}$; the flux extrapolation to the 0.2--12 keV energy band gives $\sim$$1\times10^{-12}$ 
erg cm$^{-2}$ s$^{-1}$, roughly a factor of 7 below the \emph{XMM-Newton} Slew one.

Considering the star's optical class, its X-ray properties and the persistent nature of the 
IBIS detection we conclude that HIP 87368 (HD 162186) is quite unlikely to emit at soft 
gamma-ray energies, leaving XRT object \#1 as the most likely counterpart to IGR 
J17508$-$3219.

\begin{table*}
\begin{center}
\caption{Optical/IR associations with XRT candidate counterparts discussed in the text.}
\label{tab3}
\scriptsize
\begin{tabular}{llcc}
\hline
\hline
XRT source  &   Optical/IR source$^{\dagger}$         &  Distance$^{\ddagger}$  &   Magnitudes       \\         
            &                                         &  (arcsec)  &                                   \\ 
\hline 
\hline           
\multicolumn{4}{c}{\textbf{SWIFT J0800.7$-$4309}}\\
   &    &   &       \\
\#1         &  USNO--A2.0 U0450.05566363      &  2.28 &  $R=16.8$, $B=16.5$                \\
            &  2MASS J08003998$-$4311076      &  2.22 &  $J = 15.983\pm0.103$, $H = 15.732\pm0.148$, 
$K = 15.575\pm0.213$  \\
            &  allWISE J080039.96$-$431107.2  &  2.34 &  $W1 = 15.310\pm0.036$, $W2=15.168\pm0.066$, 
$W3 > 12.888$, $W4 > 9.141$    \\
  &   &   &   \\
\#2         &  USNO--A2.0 U0450.05571649      &  3.66 &  $R=17.4$, $B=17.6$                \\
            &  allWISE J080045.83$-$430939.3  &  1.86 &  $W1 = 15.103\pm0.032$, $W2=14.435\pm0.041$, 
$W3=11.871\pm0.223$, $W4 > 8.513$    \\
\hline
\multicolumn{4}{c}{\textbf{SWIFT J0924.2$-$3141}}\\
   &   &   &   \\
\#1         &  USNO--A2.0 U0525.11601717      &  1.62 &  $R=11.7$, $B=12.9$             \\
            &  2MASS J09235373$-$3141308      &  1.68 &  $J = 14.242\pm0.087$, $H = 13.515\pm0.103$, 
$K = 12.979\pm0.071$     \\
            & allWISE J092353.73$-$314130.9   &  1.68 &  $W1 = 11.998\pm0.023$, $W2=11.591\pm0.021$, 
$W3=9.162\pm0.031$, $W4=6.790\pm0.080$    \\
  &   &   &   \\
\#2         &  USNO--A2.0 U0525.11615396      &  1.32 &  $R=17.3$, $B=19.2$           \\
\hline
\multicolumn{4}{c}{\textbf{IGR J14059$-$6116}}\\
   &      &  &    \\
\#1         &  2MASS J14051441$-$6118282      &  3.78 &  $J > 15.962$, $H = 14.369\pm0.068$,
$K = 12.769\pm0.044$    \\
            & G311.6718+00.3053	 &  3.84   &  3.6$\mu$m = $11.572\pm0.058$,
4.5$\mu$m = $11.22\pm0.061$, 5.8$\mu$m =$11.038\pm0.088$, 8.0$\mu$m =$11.020\pm0.076$  \\
            &  allWISE J140514.40$-$611827.7  &  3.90 &  $W1 = 11.612\pm0.037$, $W2=11.231\pm0.046$, 
$W3 > 9.950$, $W4 > 7.585$     \\
\hline
\multicolumn{4}{c}{\textbf{IGR J15038$-$6021}}  \\
   &    &   &        \\
\#1         &  USNO--B1.0 0296--0547603       &  3.18 &  $R1=17.77$, $R2=19.04$               \\
            &  USNO--B1.0 0296--0547602       &  3.36 &  $R2=18.63$                           \\
            &  2MASS J15041611$-$6021225      &  1.32 &  $J > 15.824$, $H > 14.594$, 
$K = 14.629\pm0.106$        \\ 
\hline
\multicolumn{4}{c}{\textbf{IGR J16447$-$5138}}  \\
   &   &   &   \\
\#1         &  USNO--B1.0 0384--0622801       &  3.18  &  $R2 = 15.96$, $B2 = 16.14$   \\
            &  2MASS J16443324$-$5134131      &  3.42  &  $J = 14.155\pm0.062$, $H = 13.152\pm0.065$,
$K = 12.606\pm0.057$        \\
            &  allWISE J164433.25$-$513413.2  &  3.48  &  $W1 = 11.625\pm0.025$, $W2=11.484\pm0.021$,
$W3=8.995\pm0.036$, $W4=6.814\pm0.058$    \\
\hline
\multicolumn{4}{c}{\textbf{IGR J17508$-$3219}}  \\
   &   &   &   \\
\#1         &  USNO--B1.0 0576--0768321       &  4.50  &  $R2 = 75.83$  \\
\#1         &  2MASS J17505271$-$3219488      &  4.26  &  $J > 14.424$, $H > 13.076$, $K = 13.179\pm0.080$ \\ 
            &  G357.6922$-$02.7203            &  2.40  &  3.6$\mu$m = $12.943\pm0.065$,
4.5$\mu$m = $12.916\pm0.099$, 5.8$\mu$m =$12.162\pm0.191$   \\
\hline
\multicolumn{4}{c}{\textbf{IGR J18007$-$4146}}  \\
   &   &   &   \\
\#1         &  USNO--B1.0 0482--0651082        &  1.38  &  $B2 = 15.36$    \\
            &  USNO--B1.0 0482--0651086        &  1.92  &  $R1 = 15.07$, $R2 = 15.28$, $B2 = 15.49$  \\
            &  2MASS J18004247$-$4146466       &  2.16  &  $J = 16.408\pm0.141$, $H = 15.800\pm0.198$,
$K = 15.240\pm0.185$        \\
            &  2MASS J18004270$-$4146503       &  2.76  &  $J = 15.519\pm0.071$, $H = 15.429\pm0.131$,
$K = 15.085\pm0.152$        \\
            &  WISE J180042.63$-$414648.9      &  1.50  &  $W1 = 14.712\pm0.047$, $W2=14.930\pm0.113$,
$W3 > 12.785$, $W4 > 9.183$    \\
\hline
\multicolumn{4}{c}{\textbf{IGR J18074$+$3827}}  \\
   &   &   &   \\
\#1         &  USNO--A2.0 U1275.09785770      &  5.52  &  $R= 12.9$, $B=14.2$        \\
            &  2MASS J18075291$+$3822384      &  4.74  &  $J = 11.006\pm0.020$, $H = 10.465\pm0.024$,
$K = 10.340\pm0.018$        \\
            &  allWISE J180752.92$+$382238.5  &  4.50  &  $W1 = 10.348\pm0.023$, $W2=10.371\pm0.020$,
$W3=10.309\pm0.058$, $W4 > 8.698$    \\	
\hline           
\multicolumn{4}{c}{\textbf{XMMSL1 J182831.8$-$022901}}  \\
   &   &   &   \\
\#1         &   UGPS J182831.01$-$022906.6         &  1.2  &  $K = 17.077\pm0.086$  \\
            &   UGPS J182831.03$-$022908.3         &  1.86 &  $K = 16.234\pm0.040$   \\
           \hline
\multicolumn{4}{c}{\textbf{SWIFT J1839.1$-$5717}}  \\
   &        &    &      \\
\#1         &  allWISE J183905.95$-$571505.1  &  3.42  &  $W1 = 13.046\pm0.024$, $W2=11.209\pm0.021$
$W3=7.740\pm0.018$, $W4=5.848\pm0.045$  \\
\hline
\multicolumn{4}{c}{\textbf{IGR J20310$+$3835}}   \\
   &    &   &       \\
\#1         &  UGPS J203055.29$+$383347.1     &  3.06  &  $J = 18.826\pm0.050$, $H = 17.297\pm0.022$, 
$K=16.544\pm0.040$   \\
            &  UGPS J203055.30$+$383347.2     &  3.12  &  $J = 18.710\pm0.048$, $H = 17.347\pm0.024$, 
$K=16.463\pm0.040$    \\
\hline
\multicolumn{4}{c}{\textbf{1SWXRT J230642.8$+$550817}}  \\
   &    &   &      \\
\#1         &  USNO--A2.0 U1425.14606199      &  3.66  &  $R=16.6$, $B=17.3$   \\
            &  2MASS J23064269$+$5508200      &  3.36  &  $J = 15.857\pm0.071$, $H = 15.688\pm0.161$,
$K = 15.392\pm0.181$     \\
            &  allWISE J230642.67$+$550820.1  &  3.18  &  $W1 = 15.118\pm0.036$, $W2=15.135\pm0.068$, 
$W3 > 13.115$, $W4 > 9.368$    \\
\hline
\hline
\end{tabular}
\begin{list}{}{}
\item $^{\dagger}$ The catalogs are the Two Micron All Sky Survey (2MASS, Skrutskie et al. 2006),
the United States Naval Observatory (USNO--B1.0 and USNO--A2.0, Monet 1998,2003),
the Wide-field Infrared Survey Explorer all sky survey (WISE, Wright et al. 2010) or, if available,
allWISE (available at:
\url{http://vizier.u-strasbg.fr/viz-bin/VizieR?-source=II/328)}, 
the Galactic Legacy Infrared Mid-Plane Survey Extraordinaire (GLIMPSE, Churchwell et al. 2009),
and the UKIRT Infrared Deep Sky Survey (UKIDSS) Galactic Plane Survey (UGPS, Lucas et al. 2008);
\item $^{\ddagger}$ Angular distance from the XRT cetroid.
\end{list}
\end{center}
\end{table*}

\subsection{\bfseries{IGR J18007$-$4146}\\ 
(\textit{Detected as a persistent source})}

\begin{figure}
\includegraphics[width=1.0\linewidth]{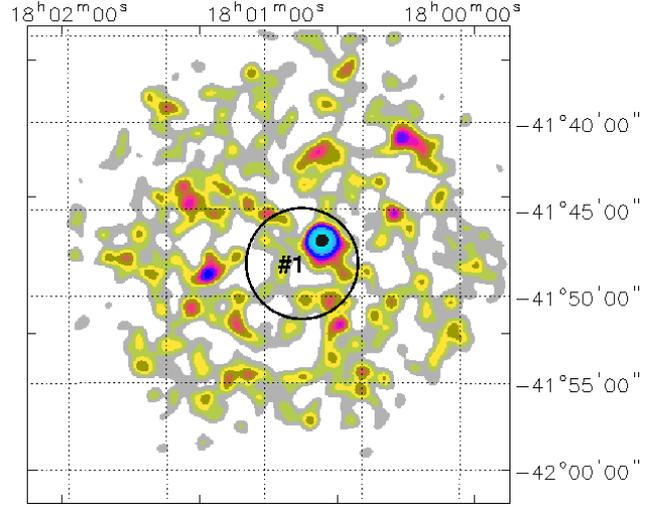}
\caption{XRT 0.3--10 keV image of the region surrounding IGR J18007$-$4146.
XRT detects only one source that is located
within the 90\% IBIS positional uncertainty (black circle).}
\label{fig14}
\end{figure}

For this \emph{INTEGRAL} emitter, only one X-ray source is detected within the 90\% IBIS 
positional uncertainty (see Figure~\ref{fig14} and Table~\ref{tab2}). It is detected at 
9.5$\sigma$ and 5.5$\sigma$ c.l. in the 0.3--10 and 3--10 keV energy range, respectively.

Within the XRT error box we also found an \emph{XMM-Newton} Slew object (XMMSL1 
J180042.8$-$414651) that is detected at 2.7$\sigma$ c.l. with a 0.2--12 keV flux of 
1.78$\times10^{-12}$ erg cm$^{-2}$ s$^{-1}$. Another \emph{XMM-Newton} Slew object (XMMSL1 
J180042.8$-$414656) was found at 8.4 arcsec from the XRT centroid. It is detected at 
2.6$\sigma$ c.l. with a 0.2--12 keV flux of 2.79$\times10^{-12}$ erg cm$^{-2}$ s$^{-1}$. The 
distance between the two \emph{XMM-Newton} Slew detections is only 6 arcsec, whereas their 
respective error radii are 3.6 and 4.1 arcsec: this strongly suggests that they are probably 
the same source seen in different periods (observing times 2010--10--06 and 2013--03--08 for 
the first and second \emph{XMM-Newton} Slew objects, respectively) and both are associated 
with the single XRT detection. If the two \emph{XMM-Newton} Slew objects are the same object, 
then the observed fluxes indicate that the source may be variable on yearly timescales.

\begin{figure}
\includegraphics[width=1.1\linewidth]{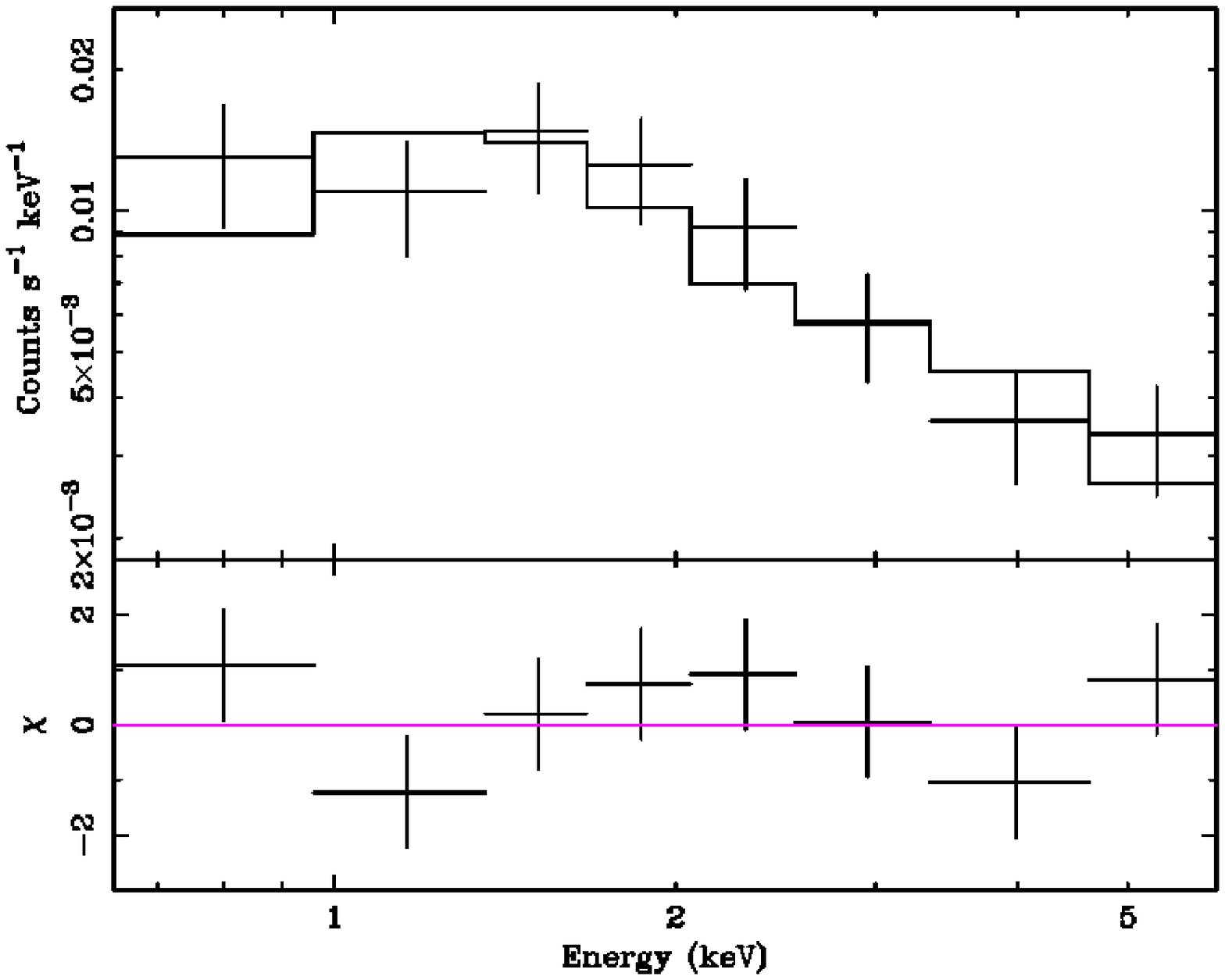}
\caption{XRT spectrum of IGR J18007$-$4146 fitted with our basic model 
(\emph{upper panel}); residuals to this model are in units of $\sigma$ (\emph{lower panel}).}
\label{fig15}
\end{figure}

Our XRT baseline model provides a flat photon index ($\Gamma \sim 1$) and a 2--10 keV flux of 
$2.7\times10^{-12}$ erg cm$^{-2}$ s$^{-1}$ (see Figure~\ref{fig15} and Table~\ref{tab4}); the 
extrapolation of the XRT flux to the 0.2--12 keV energy range yields a flux of 
$3.8\times10^{-12}$ erg cm$^{-2}$ s$^{-1}$, higher than those shown by the \emph{XMM-Newton} 
Slew detections. 

Within the XRT positional uncertainty we find two possible counterparts (see 
Table~\ref{tab2}); optical spectroscopy of both of them is necessary to disintangle which of the two 
is the real counterpart and assess its nature.

\subsection{\bfseries{IGR J18074$+$3827}\\
(\textit{Detected in a 447.2-day outburst from MJD = 53275.0})}

In this case the source detection by IBIS is optimised over a period of around 15 months 
starting from September 26, 2004, while the only XRT pointing was carried out much later, at 
the beginning of June 2016.

\begin{figure}
\includegraphics[width=1.0\linewidth]{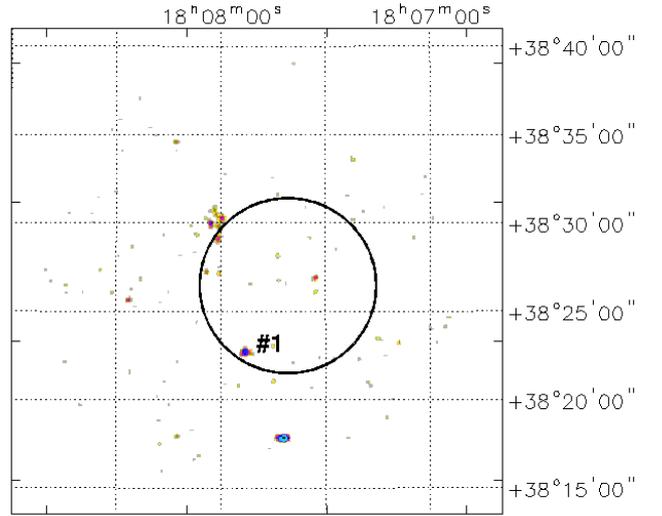}
\caption{XRT 0.3--10 keV image of the region surrounding IGR J18074$+$3827.
Only one source is found by XRT within the
90\% IBIS positional uncertainty (black circle).}
\label{fig16}
\end{figure}

In the region surrounding IGR J18074$+$3827 there is only one X-ray source that lies within 
the 90\% IBIS positional uncertainty (see Figure~\ref{fig16}). It is revealed at 3.3$\sigma$ 
in the 0.3--10 keV energy range, but not above 3 keV. The XRT error circle is compatible with 
that of an \emph{XMM-Newton} Slew source (XMMSL1 J180752.6$+$382240, error radius of 3.4 
arcsec), which is detected at 2.1$\sigma$ c.l. with a 0.2--12 keV flux of 1.55$\times10^{-12}$ 
erg cm$^{-2}$ s$^{-1}$.

Given the poor quality of the XRT data, we can only infer a 2--10 keV flux of 
$\sim$$1\times10^{-13}$ erg cm$^{-2}$ s$^{-1}$, by freezing the photon index to 1.8; 
comparison with \emph{XMM-Newton} Slew detection indicates strong X-ray variability. This 
suggests that XRT pointed at the source during a period of quite low flux, while 
\emph{INTEGRAL} and \emph{XMM-Newton} observed it during a much brighter flux state. Within 
the XRT positional uncertainty we find a single optical/infrared counterpart that is also 
reported as a bright UV source (GALEX J180752.91$+$382238.9 with Near and Far UV magnitudes of 
17.9 and 17.6, respectively). The UV detection and the source location, well above the 
Galactic plane ($b= \sim+24.6^{\circ}$), argue for an extragalactic nature, but unfortunately 
there is no radio counterpart associated with the optical-UV/IR counterpart nor are its WISE 
colours compatible with an AGN nature for the source. Using the 2MASS magnitudes to 
compute the free reddening parameter Q = (J--H)--1.7(H--Ks) to create a Q/Ks diagram 
(Neguerela \& Schurch 2007), we find that the source falls in the region of late type stars 
(Reig \& Milonaki 2016) casting doubts on its detection above 20 keV. Furthermore, the lack of 
emission above 3 keV and the variability seen mostly at X-ray energies further complicate the 
issue and provide indication that the XRT/\emph{XMM-Newton} Slew source may not be the correct 
association to the \emph{INTEGRAL} object.

Clearly, optical spectroscopy of the only X-ray counterpart found together with further 
observations at time of strong X-ray emission can shed light and eventually help to classify 
this source.

\subsection{\bfseries{XMMSL1 J182831.8$-$022901 (also 3PBC J1828.7$-$0227)}\\
(\textit{Detected as a persistent source})}

\begin{figure}
\includegraphics[width=1.0\linewidth]{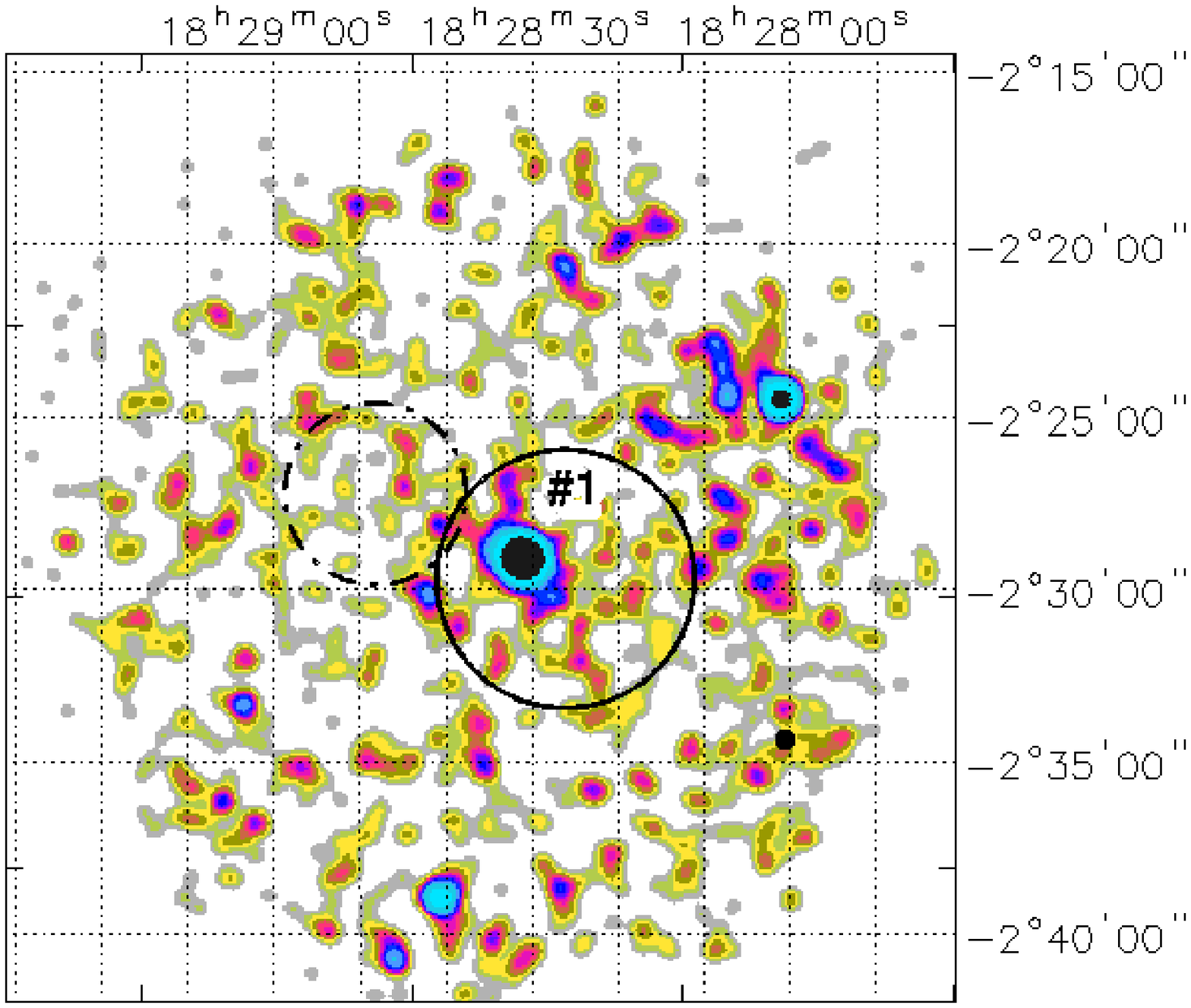}
\caption{XRT 0.3--10 keV image of the region surrounding XMMSL1 J182831.8$-$022.
The only X-ray detection is located
inside the 90\% IBIS positional uncertainty (black circle). The 90\% BAT (black-dashed-dotted circle) 
positional uncertainty partially overlaps the IBIS one, but does not include source \#1.}
\label{fig17}
\end{figure}

This source is also reported in the Palermo 66-month \emph{Swift}/BAT hard X-ray catalogue as 
3PBC J1828.7$-$0227\footnote{Available at:\\ 
\url{http://bat.ifc.inaf.it/bat_catalog_web/66m_bat_catalog.html.}}. XRT detects only one 
source which lies within the 90\% IBIS error circle, but just outside the 90\% BAT positional 
uncertainty (see Figure~\ref{fig17}). It is seen at 11.6$\sigma$ and 8.9$\sigma$ c.l. in the 
0.3--10 and 3--10 keV, respectively (see Table~\ref{tab2}). The XRT position is 
compatible within respective uncertainties with the location of the \emph{XMM-Newton} Slew 
detections XMMSL1 J182831.8$-$022901 and XMMSL1 J182831.4$-$022914, which are detected at 2.7 
and 2.3 $\sigma$ c.l. in the 0.2--12 keV energy range with a flux of $1.5\times10^{-12}$ erg 
cm$^{-2}$ s$^{-1}$ and $2.3\times10^{-12}$ erg cm$^{-2}$ s$^{-1}$ respectively. Both 
detections correspond to a single source seen by \emph{XMM-Newton} at different epochs 
(September 23 and October 12, 2012) and possibly varying over time.

This X-ray source has only a couple of possible IR counterparts (see Table~\ref{tab3}).

\begin{figure}
\includegraphics[width=1.1\linewidth]{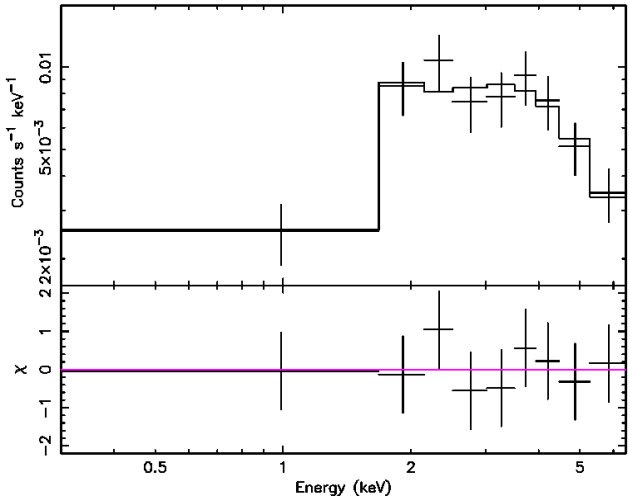}
\caption{XRT spectrum of XMMSL1 J182831.8$-$022 with our basic model plus 
intrinsic absorption (\emph{upper panel}); 
residuals to this model are in units of $\sigma$ (\emph{lower panel}).}
\label{fig18}
\end{figure}

Our basic power law model does not provide a good fit to the XRT data, which require 
additional intrinsic absorption ($N_{\rm{H(intr)}} \sim 1.5\times10^{22}$ cm$^{-2}$). The 
photon index turns out to be around 1.2 and the 2--10 keV (0.2--12 keV) flux is $\sim$$4.6 
(5.9)\times 10^{-12}$ erg cm$^{-2}$ s$^{-1}$ (see Table~\ref{tab4} and Figure~\ref{fig18}), 
which suggests some variability (by a factor of 2.6) if compared to the \emph{XMM--Newton} 
Slew ones.

The location of XMMSL1 J182831.8$-$022901 on the Galactic plane ($b = 3.95^{\circ}$) indicates 
that we may be dealing with either a Galactic source (i.e. some type of X-ray binary) or an 
AGN hidden behind the Galactic plane. Only IR spectroscopy of the likely counterparts can 
disintangle which is the correct association and eventually unveils its nature.

\begin{table*}
\begin{center}
\caption{\emph{Swift}/XRT spectral analysis results of the averaged spectra. Frozen parameters are
written in square brackets; errors are given at the 90\% confidence level.}
\label{tab4}
\begin{tabular}{lcccccc}
\hline
\hline
 Source & $N_{\rm H(Gal)}$ & $N_{\rm H(int)}$ & $\Gamma$ & $\chi^{2}/d.o.f.$ & $C-stat/d.o.f.$ &  $F_{\rm (2-10~keV)}$ \\
  & ($10^{22}$ cm$^{-2}$) &  ($10^{22}$ cm$^{-2}$)   &  &  &  & ($10^{-11}$ erg cm$^{-2}$ s$^{-1}$) \\
\hline
\hline
\multicolumn{6}{c}{\textbf{SWIFT J0800.7$-$4309}}\\
   &   &   & & &  \\
\#1        &  0.353  &         --               &   $0.83\pm0.11$   &    43.3/35 &  -- &  $0.50\pm0.02$ \\
   &   &   &   &  &  \\
\#2        &  0.355  &         --               &    $1.22\pm0.56$ &  --        & 46.5/46  & $0.021\pm0.004$ \\
\hline
\multicolumn{6}{c}{\textbf{SWIFT J0924.2$-$3141}}\\
   &   &   & & & \\
\#1$^{a}$  &  0.133  &  $81.8^{+56.8}_{-38.5}$  &        [1.8]        & --  & 86.8/71    &  $0.17\pm0.06$ \\
&   &   &   & &    \\
\#2$^{b}$  &  0.132  &         --               &   $1.69\pm0.11$     & 221.3/213 &   --   & $19.5\pm0.26$  \\           
\hline
\multicolumn{6}{c}{\textbf{IGR J14059$-$6116}}\\
   &   &   & & &   \\
\#1        &  1.84   &            --            &        [1.8]        & --   &  52.4/38  & $0.03\pm0.005$  \\
\hline
\multicolumn{6}{c}{\textbf{1RXSJ 145959.4$+$120124}}\\
   &   &   & & &  \\
\#1        & 0.0232  &             --           &  $1.89\pm0.23$  & 12.0/9   &   --         & $0.24\pm0.02$ \\
\hline
\multicolumn{6}{c}{\textbf{IGR J15038$-$6021}}\\
   &   &   & & &  \\
\#1        & 1.26    &             --           &        [1.8]          & --   & 4.0/7        & $0.16\pm0.03$   \\
\hline
\multicolumn{6}{c}{\textbf{IGR J16447$-$5138}}\\
   &   &   &  & &  \\
\#1        & 0.387   &     $<$ 0.6              &    $1.96^{+0.58}_{-0.48}$  & 10.7/9  &  --  &  $0.39\pm0.03$ \\
\hline
\multicolumn{6}{c}{\textbf{IGR J17508$-$3219}}\\
   &   &   &    & & \\
\#1$^{c}$        & 0.452   &            --            &    $0.64^{+0.71}_{-0.78}$ & --  & 27.5/25   & $0.08\pm0.02$  \\
&  &  &  & &   \\
           & 0.452   & $1.15^{+1.55}_{-0.86}$         &       [1.8]       &  --   & 28.5/25      & $0.07\pm0.02$   \\
\hline
\multicolumn{6}{c}{\textbf{IGR J18007$-$4146}}\\
   &   &   & & &   \\
\#1        & 0.120   &            --            &    $1.02\pm0.30$     & 5.9/6     &  -- &  $0.27\pm0.03$  \\
\hline
\multicolumn{6}{c}{\textbf{IGR J18074$+$3827}}\\
   &   &   &  & & \\
\#1        & 0.0273   &               --        &         [1.8]    &  --    & 18.5/18       &  $0.010\pm0.002$ \\
\hline
\multicolumn{6}{c}{\textbf{XMMSL1 J182831.8$-$02290}}\\
   &   &   &  &  & \\
\#1        & 0.512   &   $1.53^{+1.34}_{-0.98}$ &  $1.21^{+0.75}_{-0.66}$   &  5.3/6  &  --   &  $0.46\pm0.03$ \\
\hline
\multicolumn{6}{c}{\textbf{SWIFT J1839.1$-$5717}}\\
   &   &   & & &  \\
\#1        & 0.0726  & $2.06^{+0.36}_{-0.33}$   &    $1.57\pm0.21$ &    71.7/63  &   --   &  $0.84\pm0.02$ \\
\hline
\multicolumn{6}{c}{\textbf{IGR J20310$+$3835}}\\
   &   &   & &  &  \\
\#1$^{c}$  & 1.04    &        --                &  $0.14^{+0.68}_{-0.75}$ &  --    &  44.2/58    &  $0.18\pm0.02$ \\
   &   &   &   &  &  \\
           & 1.04    & $5.58^{+7.62}_{-3.56}$   &       [1.8]             & --     &  45.8/57    &  $0.12\pm0.02$   \\
\hline
\multicolumn{6}{c}{\textbf{1SWXRT J230642.8$+$550817}}\\
   &   &   &  &  & \\
\#1        & 0.311   &           --             &  $1.00\pm0.47$  &   --    & 44.3/52   &    $0.67\pm0.09$  \\
\hline
\hline
\end{tabular}
\begin{list}{}{}
$^{a}$ In this case, the best-fit model requires a second power law component, having the same photon
index of the primary absorbed power law, and passing only through the Galactic column density;\\
$^{b}$ In this case, the best-fit model includes a black-body component ($kT = 1.09\pm0.09$ keV)
to account for the excess observed below 2 keV;\\
$^{c}$ For this source, we report the results of the spectral analysis obtained both by leaving the photon 
index to vary and by freezing it to 1.8 (see text).
\end{list}
\end{center}
\end{table*}

\subsection{\bfseries{SWIFT J1839.1$-$5717}\\
(\textit{Detected as a persistent source})} 

\begin{figure}
\includegraphics[width=1.0\linewidth]{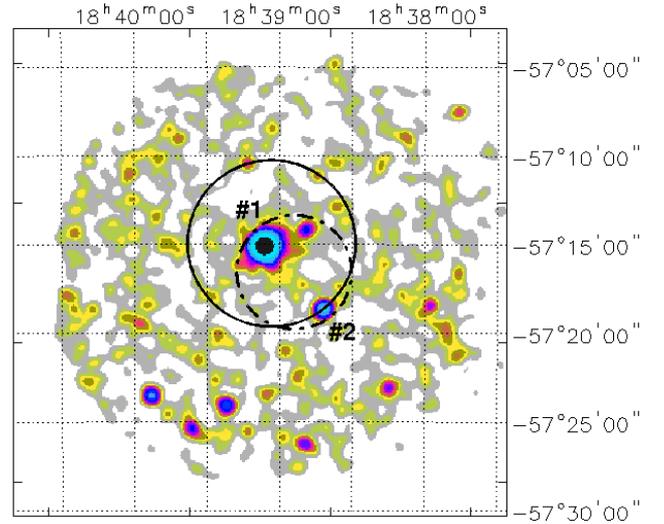}
\caption{XRT 0.3--10 keV image of the region surrounding SWIFT J1839.1$-$5717.
Two sources are detected by XRT: source \#1 is located within the 90\% IBIS error circle 
(black circle),
while source \#2 lies at its border. Both sources are instead located within the 
90\% BAT positional uncertainty
(black-dashed-dotted circle).}
\label{fig19}
\end{figure}

This source is also listed in the 70-month \emph{Swift}/BAT catalogue (Baumgartner et al. 
2013) and it is one of the few objects in our sample located off the Galactic plane ($b= 
-20.95^{\circ}$). XRT follow-up observations indicate the presence of two X-ray sources whose 
positions are compatible with the 90\% IBIS/BAT error circles (see Figure~\ref{fig19} and 
Table~\ref{tab2}).

Source \#1 is the brightest of the two X-ray detections (around 32.5$\sigma$ and 23.8$\sigma$ 
c.l. in the 0.3--10 and 3--10 keV, respectively) and also the only one still detected above 3 
keV; it is positionally compatible with the allWISE source listed in Table~\ref{tab2}.

\begin{figure}
\includegraphics[width=1.1\linewidth]{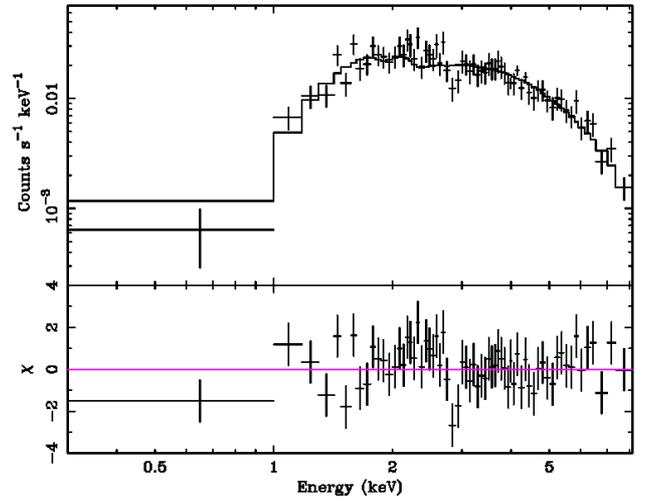}
\caption{XRT spectrum of SWIFT J1839.1$-$717 fitted with our basic model plus intrinsic absorption
(\emph{upper panel}); 
residuals to this model are in units of $\sigma$ (\emph{lower panel}).}
\label{fig20}
\end{figure}

Its X-ray spectrum requires intrinsic absorption ($N_{\rm{H(int)}} \sim 2\times10^{22}$ 
cm$^{-2}$) and shows a photon index $\Gamma \sim 1.6$ and 2--10 keV flux of 
$\sim$$4\times10^{-12}$ erg cm$^{-2}$ s$^{-1}$ (see Table~\ref{tab4} and Figure~\ref{fig20}).

Thus, the XRT data univocally bring us to consider the association of source \#1 with SWIFT 
J1839.1$-$5717 highly likely. Not only is it the brightest and hardest object detected in 
X-rays, but it is also listed in the WISE AGN catalogue by Secrest et al. (2015) since its 
WISE colours ($W1 -W2 = 1.84$ and $W2 - W3 = 3.47$) are typical of an IR-selected AGN. 
Furthermore, its location off the Galactic plane strengthens its extragalactic nature. Last 
but not least, its spectral behaviour suggests a type 2 AGN, i.e. the X-ray spectrum requires 
intrinsic absorption in excess to the Galactic one.

\subsection{\bfseries{IGR J20310$+$3835 (also 3PBC J2030.8$+$3833)}\\ 
(\textit{Detected as a persistent source})}

This source is listed in the Palermo 66 month \emph{Swift}/BAT hard X-ray catalogue
as 3PBC J2030.8$+$3833.

\begin{figure}
\includegraphics[width=1.0\linewidth]{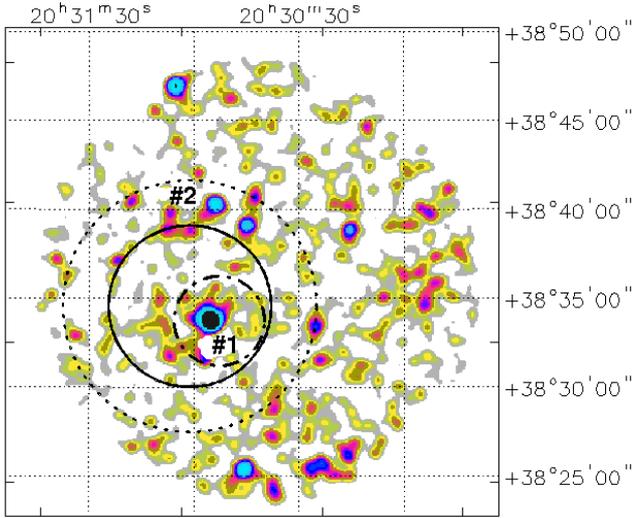}
\caption{XRT 0.3--10 keV image of the region surrounding IGR J20310$+$3835.
The two XRT detections (sources \#1 and \#2) are located within the 
within the 90\% (black circle) and 99\% (black-dotted circle) IBIS positional uncertainties,
respectively. Source \#1 also lies within the 90\% BAT error circle (black-dashed-dotted circle).}
\label{fig21}
\end{figure}

XRT detects two sources whose positions are compatible with either the 90\% or 99\% IBIS error 
circle (see Figure~\ref{fig21} and see Table~\ref{tab2}), but only one is likely associated 
with the IBIS/BAT emitter.

Source \#1, which is detected at 7.4$\sigma$ c.l. in the range 0.3--10 keV, is the only one 
detected above 3 keV (6.2$\sigma$ c.l. in the range 3--10 keV) and hence the hardest of the 
two; it is also the only source compatible with the BAT positional uncertainty. Only two IR 
counterparts were found for this XRT object (see Table~\ref{tab2}).

By fitting the XRT data with our basic model we find a flat photon index ($\Gamma \sim 
0.1$) and a 2--10 keV flux of $\sim$$2\times10^{-12}$ erg cm$^{-2}$ s$^{-1}$. If we freeze the 
photon index to 1.8, the data require an intrinsic column density $N_{\rm{H(intr)}} \sim 
6\times10^{22}$ cm$^{-2}$; in this case the 2--10 keV flux is around $1\times10^{-12}$ erg 
cm$^{-2}$ s$^{-1}$ (see Table~\ref{tab4}).

These findings, combined with the location of IGR J20310$+$3835 on the Galactic plane ($b = 
-0.49^{\circ}$), suggest again that we may be dealing with either a Galactic source or an 
absorbed AGN hidden behind the Galactic plane. Only IR spectroscopy of the likely counterparts 
can discriminate between these two options.

\subsection{\bfseries{1SWXRT J230642.8$+$550817}\\
(\textit{Detected as a persistent source})}

\begin{figure}
\includegraphics[width=1.0\linewidth]{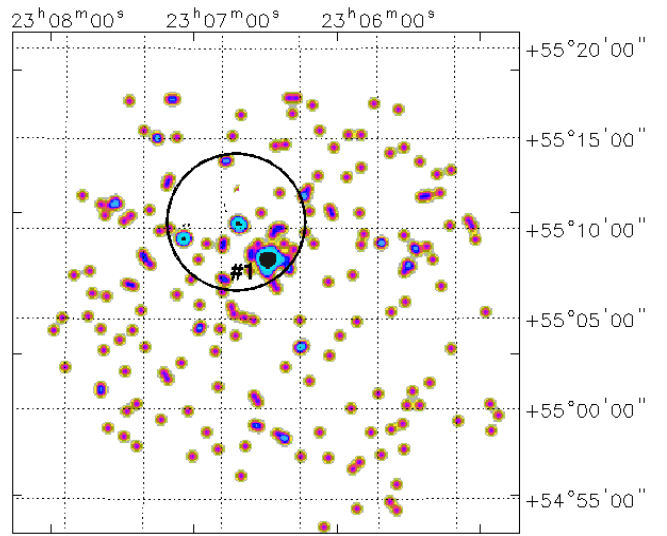}
\caption{XRT 0.3--10 keV image of the region surrounding 1SWXRT J230642.8$+$550817.
Only one X-ray source is detected by XRT within the 90\% IBIS positional uncertainty (black circle).}
\label{fig22}
\end{figure}

As shown in Figure~\ref{fig22} only one X-ray source is detected by XRT within the 90\% IBIS 
positional uncertainty at 7.4$\sigma$ and 4.4$\sigma$ c.l. in the 0.3--10 and 3--10 keV energy 
band, respectively. It has a single optical/infrared counterpart as reported in 
Table~\ref{tab2}. This source is also listed as an H$_{\alpha}$ emission line object in the 
INT/WFC Photometric H$_{\alpha}$ Survey (IPHAS, Witham et al. 2008). Indeed, it has recently 
been classified as a CV by Masetti et al. (in preparation), who will provide details on the 
optical spectrum in a forthcoming paper.

The XRT spectrum shows a flat photon index ($\Gamma \sim 1$) and a 2--10 keV flux of 
$\sim$$7\times10^{-12}$ erg cm$^{-2}$ s$^{-1}$ (see Table~\ref{tab4}). The source properties 
and its optical classification argue strongly in favour of its association with the newly 
reported IBIS object.

\section{Summary and conclusions}

In this paper, we report on the results of a work aimed at exploiting \emph{Swift}/XRT 
archival data to search for candidate counterparts to a set of still unidentified sources 
listed in the latest \emph{INTEGRAL}/IBIS survey (Bird et al. 2016). Only in one case, we were 
not able to provide an XRT association, although a possible X-ray counterpart was reported in 
the \emph{XMM-Newton} Slew survey; in all other cases one or two likely associations were 
found. The more accurate position provided by XRT enabled us to pinpoint the optical, IR, and 
UV counterpart for most of these candidate associations. Moreover, for the brightest objects, 
i.e. those having a signal-to-noise ratio good enough to allow a reliable spectral analysis, 
we also characterised the X-ray spectrum, while for the fainter sources only a flux estimate 
in the 2--10 or 0.2--12 keV energy range is reported. When more than one XRT pointing were 
available and/or an association/s with \emph{XMM-Newton} Slew survey were found, we also 
explored flux varibility in the X-ray band. All the information gathered helped us to propose, 
for each IBIS source, the most likely counterpart and discuss its nature. For SWIFT 
J0924.2$-$3142 we found that the high-energy emission, although most likely due to the 
contribution of two objects (a Seyfert 2 and a bright unclassified source) up to 15 keV, it is 
related only to the AGN above these energies. Optical follow-up observations of the proposed 
counterpart to SWIFT J0800.7$-$4309 and 1SWXRT J230642.8$+$550817 has led Rojas et al. (2016) 
and Masetti et al. (in preparation) to classify both of them as CVs. The properties found for 
one of the few IBIS objects detected off the Galactic plane (SWIFT J1839.1$-$5717) suggest it 
is most likely a type 2 AGN. Finally, we note that IGR J14059$-$6116 is likely associated with 
a GeV source (3FGL J1405.4$-$6119). In all other cases, follow-up optical/IR observations are 
necessary to help to classify the proposed counterparts and assess their ultimate nature.

As a final remark, we note that the results of this work confirms the key role played by 
follow-up observations with current X-ray telescopes and the importance of multi-waveband 
studies, in particular optical/infrared spectroscopy.

\section*{Acknowledgments}
We thank the anonymous referee for useful comments/suggestions that help to improve 
the quality of the paper. This research has made use of data obtained from the SIMBAD 
database operated at CDS, Strasbourg, France; from the High Energy Astrophysics Science 
Archive Research Center (HEASARC), provided by NASA's Goddard Space Flight Center; from the 
NASA/IPAC Extragalactic Database (NED); and from the Palermo BAT catalogue and database opened 
at INAF--IASF Palermo. The authors also acknowledge the use of public data from the 
\emph{Swift} data archive.
The authors acknowledge financial support from ASI under contract \emph{INTEGRAL} ASI 
2013--025--R.0.

\end{document}